# Photoengineering the Magnon Spectrum in an Insulating Antiferromagnet

V. Radovskaia[1]*, R. Andrei[2], J.R. Hortensius[3,4], R.V. Mikhaylovskiy[5], R. Citro[6], S. Chattopadhyay[2,7], M.X. Na[1], B.A. Ivanov[1,8], E. Demler[2], A.V. Kimel[1], A.D. Caviglia[9], and D. Afanasiev[1]*

[1] *Radboud University, Institute for Molecules and Materials, 6525 AJ Nijmegen, Netherlands*

[2] *Institute for Theoretical Physics, ETH Zurich, 8093 Zurich, Switzerland*

[3] *Kavli Institute of Nanoscience, Delft University of Technology, P.O. Box 5046, 2600 GA Delft, Netherlands*

[4] *Electromagnetic Signatures and Propagation, TNO, 2597 AK The Hague, The Netherlands*

[5] *Department of Physics, Lancaster University, Bailrigg, Lancaster LA1 4YW, United Kingdom*

[6] *Dipartimento di Fisica "E. R. Caianiello", Università degli Studi di Salerno and CNR-SPIN, Via Giovanni Paolo II, I-84084 Fisciano (Sa), Italy*

[7] *Lyman Laboratory, Department of Physics, Harvard University, Cambridge, MA 02138, USA*

[8] *Institute of Magnetism, National Academy of Sciences of Ukraine, Kiev, 03142 Ukraine*

[9] *Department of Quantum Matter Physics, University of Geneva, CH-1211 Geneva 4, Switzerland*

*Correspondence to: viktoriia.radovskaia@ru.nl, d.afanasiev@science.ru.nl

**Femtosecond optical pulses have opened a new frontier in ultrafast dynamics, enabling direct access to fundamental interactions in quantum materials. In antiferromagnets (AFMs), where the fundamental quantum mechanical exchange interaction governs spin dynamics, this access is especially compelling, enabling the excitation of magnons – collective spin-wave modes – that naturally reach terahertz (THz) frequencies and supersonic velocities. Femtosecond optical pulses provided a route to coherently excite such magnons across the entire Brillouin zone. Controlling their spectral properties – such as the magnon gap and dispersion – represents the next monumental step, enabling dynamic tuning of group velocities, coherence, and interaction pathways. Yet, achieving this remains a challenge, requiring ultrafast and long-lasting manipulation of the underlying exchange interaction.**

**Here, we show that in $DyFeO_3$ – an insulating AFM with strongly coupled electronic and magnetic degrees of freedom – resonant above-bandgap optical excitation leads to a dramatic renormalization of the THz magnon spectrum, including a near-total collapse of the magnon gap. Our analysis reveals this transformation to be consistent with a transient reduction of the exchange interaction by nearly 90% in the near-surface nanoscale region. These findings establish a pathway for light-driven, nanoscale control of AFM spin dynamics, opening opportunities for reconfigurable, high-speed magnonic and spintronic applications.**

Femtosecond (*fs*) light pulses lead the exploration of spin dynamics in antiferromagnets (AFMs) – materials with antiparallel spin ordering. In response to such rapid excitation, AFMs generate magnons – collective waves of spin precession at terahertz (THz) frequencies – offering exciting opportunities for high-speed and energy-efficient spintronics and magnonics applications[1–4]. Ultrashort pulses have been shown to drive coherent spin-wave excitations across the entire Brillouin zone, including zone-center magnons[5], zone-edge two-magnon states[6], and, more recently, finite-momentum propagating modes[7–10]. Light has not only enabled the excitation of these modes but also revealed their nonlinear interactions[11–14], leading to a redistribution of energy and momentum across the entire magnon spectrum. Ultrafast photoengineering of the THz magnon spectrum is the next monumental challenge enabling precise control over magnon propagation and interactions on ultrafast timescales[15,16]. Achieving this crucially requires direct control of the most fundamental interaction in AFMs — the exchange interaction — which not only drives AFM magnon frequencies into the THz range but also establishes the quasi-linear relativistic dispersion relation between their frequency and momentum, thus enabling dispersionless propagation of AFM magnons at supersonic velocities[1,2,4].

Ultrafast optical control of exchange interactions has been widely studied in strongly correlated AFM systems, such as Mott-Hubbard and charge-transfer (*CT*) insulators. In these systems, local electron-electron correlations not only determine the nature of the insulating bandgap but also govern the strength of the AFM exchange. Current approaches to optical control the exchange interaction rely on tuning electron wavefunction overlap – i.e. hopping amplitudes – by below- or above-bandgap optical excitation[17–21]. While below-bandgap excitation enables reversible control of exchange interaction, the effects are short-lived, typically limited to the optical pulse duration[17,20,22]. Recent experiments also suggest that such modifications are on the order of 1% and are primarily limited to impulsive excitations of coherent spin precession[18,21,23]. In contrast, near or above-bandgap optical excitation creates a nonequilibrium distribution of photoexcited charge carriers. Theory predicts that in the limit of strong electronic correlations, these carriers may relax into a nonthermal "hidden" state, transiently modifying exchange interactions[24–27]. However no direct impact on the coherent spin-wave spectrum has yet been observed; instead, the primary effect remains the ultrafast melting of AFM order, driven by the collapse of local magnetic moments[28–32].

In this work, we investigate the impact of *fs* laser excitation on the magnon dynamics in the *CT* insulating AFM oxide dysprosium orthoferrite ($DyFeO_3$), known for strong optomagnetic effects[33,34] and electronic correlations[35,36]. Using time- and momentum-resolved magnon spectroscopy, we show that at the material's surface, above-bandgap excitation generates coherent AFM spin-waves with distinct spectral signatures absent in equilibrium. Most notably, we observe the emergence of a quasi-continuous spectral distribution of magnons at reduced frequencies/energies, resulting from a near-total collapse of the equilibrium magnon energy gap. This profound modification of the spin-wave spectrum leads to extraordinary large decay rate of the experimentally observed coherent spin-wave excitations. Our theoretical model, incorporating key magnetic interactions, indicates that the observed spectral renormalization is driven by a long-lived, nonthermal electronic state confined to a photoexcited near-surface nanoscale layer, where exchange interactions are dramatically altered.

**Magnon spectrum and electronic structure of AFM DyFeO₃**

DyFeO$_3$ is an AFM insulator with a high Néel temperature of $T_N \simeq 650$ K and significant electronic correlations[35,36]. Below $T_N$ the AFM hosts two distinct magnetic phases separated by a first-order phase transition at $T_M = 51$ K.[37] In the low-temperature AFM phase, the iron $Fe^{3+}$ spin ($S=5/2$) sub-lattices $S_1$ and $S_2$ are antiparallel, resulting in zero net magnetization. In the high-temperature weak ferromagnetic (WFM) phase, the spin-orbit-driven Dzyaloshinskii-Moriya (DM) interaction causes the iron spins to cant, resulting into the emergence of a net magnetization $\mathbf{M}_0$ along the $z$-axis. Our experiments focus on the WFM phase, where spin dynamics exhibit an order of magnitude longer coherence times[38], and net magnetization enable detailed tracking of AFM spin dynamics and magnon spectra.

Similar to many other AFMs, the equilibrium magnon spectrum in DyFeO$_3$ near the Brillouin zone centre can be described by a pair of magnon branches each characterized by a dispersion relation of the form[34,38]:

$$f(k) = \sqrt{f_0^2 + (V_m k)^2} \quad (1),$$

relating the magnon frequency $f$ to its momentum, characterized by wavenumber $k$. The value $V_m$ defines the relativistic group velocity of AFM magnons; the lowest frequency $f_0$ is attributed to uniform spin precession ($k = 0$) and sets the magnon energy gap $E_0 = hf_0$. The magnon gap arises from the interplay of magnetic anisotropy and exchange interaction and - in orthoferrites - is highly sensitive to temperature through the anisotropy parameter[39]. In contrast, the group velocity $V_m$, governed solely by the exchange interaction, remains nearly constant at $V_m \simeq 20$ km/s across all temperatures[39]. Both $f_0$ and $V_m$ can thus serve as an effective measure of the exchange interaction.

The valence and conduction bands of DyFeO$_3$ are primarily composed of oxygen *2p* and iron *3d* orbitals, respectively. Photoexcitation across the bandgap ($E_{CT} \simeq 2.2$ eV) constitutes an electron transfer from an oxygen *2p* orbital to an neighbouring iron *3d* orbital, see Fig. 1a. These *CT*-transitions play a pivotal role in the magnetic properties of DyFeO$_3$: in equilibrium virtual *CT*-process mediates superexchange interactions between adjacent iron ions, thereby setting the AFM order[40,41]. Resonant pumping of the *CT*-transitions has also been shown to promote the emission of coherent magnons[9,42,43]. Yet, the effect of the excitation mechanism on the exchange interaction and on the spin-wave spectrum has remained elusive.

**Anomalous magnon dynamics at $k \simeq 0$**

To start, we study the light-induced zone-centre ($k \simeq 0$) coherent magnons in the WFM phase ($T = 55$ K) and compare the spin dynamics excited by pump pulses with photon energies from far below ($h\nu = 0.165$ eV) to above the *CT*-bandgap ($h\nu = 3.1$ eV). To access their dynamics experimentally, we detect the pump-induced magneto-optical (MO) Faraday rotation $\theta_F$ of co-propagating time-delayed probe pulses, which is sensitive to the dynamics of the magnetization $\mathbf{M}$ along the $z$-axis (Extended Data Fig. 1). Measured in the transmission geometry, $\theta_F$ captures a net component of the spin dynamics spatially integrated across the entire sample and thus the photoexcited volume.

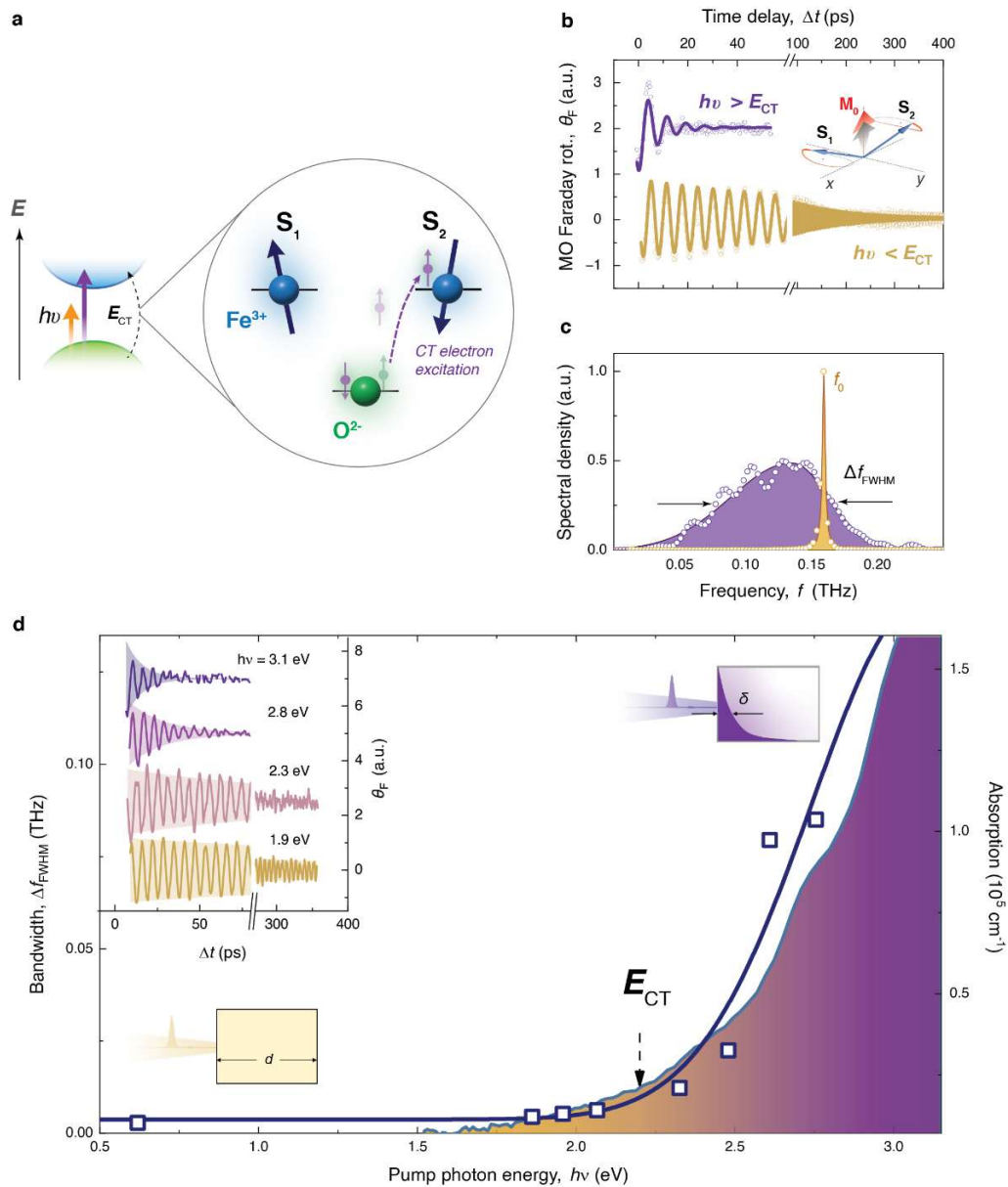

**Fig. 1| Anomalous magnon dynamics induced by selective pumping of *CT*-transitions. (a)** A schematic illustrating charge transfer (*CT*) electronic transitions between oxygen (green) and iron (blue) ions, which define the optical bandgap in $DyFeO_3$. **(b)** Time-resolved traces of the magneto-optical (MO) Faraday rotation $\theta_F$ showing the coherent magnon dynamics at $k \simeq 0$ for the optical pumping at the photon energy below ($h\nu = 0.165$ eV) and above ($h\nu = 3.1$ eV) the *CT* optical bandgap $E_{CT} = 2.2$ eV. The inset illustrates a schematic representation of the *q*-AFM spin precession mode. **(c)** Fourier transform showing the spectral density distribution of the time-traces from (b). $f_0$ denotes the frequency of $k \simeq 0$ mode excited in above *CT*-bandgap regime. **(d)** (left axis, squares) Full width at half maximum (FWHM) spectral bandwidth $\Delta f_{FWHM}$ of the pump induced magnon dynamics as a function of $h\nu$. Solid purple line is a guide-to-the-eye. (right axis) Colour-filled graph is an optical absorption profile of $DyFeO_3$. Top-left inset shows time-resolved traces of the pump-induced coherent magnon dynamics for various $h\nu$ across the *CT*-bandgap. The bottom-left and top-right insets schematically illustrate the spatial distribution of the pump-induced spin excitation through the sample volume for the regimes $h\nu < E_{CT}$ (left) and $h\nu > E_{CT}$ (right).

Figure 1b shows that both types of optical excitation drive coherent oscillatory dynamics with the frequency close to $f_0 \simeq 0.16$ THz, corresponding to the $k \simeq 0$ magnon mode of quasi-antiferromagnetic (*q*-AFM) type.[38] These uniform spin dynamics are characterized by oscillations in the canting angle of neighbouring AFM spins, modulating the magnitude of **M** without altering its orientation (inset in Fig. 1b). Despite the similar frequencies, the decay rate $\gamma$ of these *q*-AFM magnons is dramatically different: at $h\nu < E_{CT}$, $\gamma$ coincides with $\gamma_0 = 0.003$ ps$^{-1}$ for the *q*-AFM resonance[38], and at $h\nu > E_{CT}$, it yields $\gamma \simeq 0.12$ ps$^{-1}$ – nearly two orders of magnitude larger than $\gamma_0$. Fourier analysis (Fig. 1c) shows that the spectral density of the *CT*-induced magnon dynamics is significantly broader and exhibits a pronounced asymmetry. Plotting the spectral bandwidth $\Delta f_{FWHM}$ of the optically excited magnons against the pump photon energy (Fig. 1d) reveals that varying $h\nu$ has no effect on $\Delta f_{FWHM}$ below bandgap $h\nu \simeq E_{CT}$. Upon onset of *CT*-absorption, the value of $\Delta f_{FWHM}$ rises rapidly. Note that off-resonant pumping at $h\nu < E_{CT}$ exhibits negligible absorption, thereby uniformly exciting the sample across its entire thickness $d \simeq 60$ μm. In contrast, *CT*-excitation at $h\nu > E_{CT}$ is highly localized within a penetration depth $\delta \lesssim 100$ nm,[9] considerably smaller than $d$. Thus, the anomalously short-lived magnon dynamics and associated spectral broadening is related to the *CT*-excitation localized to the sample surface. Unlike off-resonant excitation, where magnon amplitude depends on pump polarization[44], resonant *CT*-excitation yields a polarization-independent response (Extended Data Fig. 2), consistent with an isotropic exchange-driven excitation mechanism[20,21].

**Photoinduced continuum of in-gap magnon states**

We further investigate how the strength of *CT*-excitation influences the dynamics of localized $k \simeq 0$ magnons in time and frequency domains. Increasing the pump fluence $F$ not only enhances the magnon amplitude but also drastically changes their temporal waveforms (Fig. 2a) and spectral profiles (Fig. 2b). In the low-fluence limit ($F = 0.6$ mJ/cm$^2$), the $k \simeq 0$ magnon spectrum shows a narrow bandwidth ($\Delta f_{FWHM} < 0.02$ THz) symmetrically centered around the equilibrium magnon gap $f_0$, similar to the behaviour observed under off-resonant pumping (Fig. 1c). The increasing pump fluence broadens the spectrum of the excited magnons, skews its spectral profile and redshifts the peak frequency $f_p$ such that $f_p < f_0$, see inset in Fig. 2b. These results suggest that near the sample surface, optical pumping of *CT*-excitations generates a broadband quasi-continuum of coherent magnon states with frequencies well below the magnon gap. Due to their non-zero spatially integrated amplitudes, these modes contribute to the MO Faraday signal, effectively mimicking $k \simeq 0$ dynamics. The observed decay thus stems from destructive interference among modes with slightly different frequencies, with coherence lost on a timescale set by the inverse bandwidth ($\Delta f_{FWHM}^{-1}$), rather then intrinsic damping.

Importantly, the in-gap magnon continuum at $k \simeq 0$ cannot be attributed to a pump-induced heating and is therefore non-thermal in nature. Although in orthoferrites the value $f_0$ is strongly temperature-dependent, in DyFeO$_3$ it does not fall below 0.14 THz[38]. In contrast, the observed in-gap magnon states have much lower frequencies and thus are not thermally accessible. Furthermore, in the studied WFM phase, the frequency $f_0$ increases with temperature (see Ref.[44] and Extended Data Fig. 3) – contrary to the fluence-driven decrease in $f_p$. The non-thermal nature of the observed magnon frequency changes, combined with the polarization-independent excitation mechanism, supports our hypothesis that *CT*-excitation in DyFeO$_3$ renormalizes the exchange interaction governing the spin-wave spectrum at $k \simeq 0$.

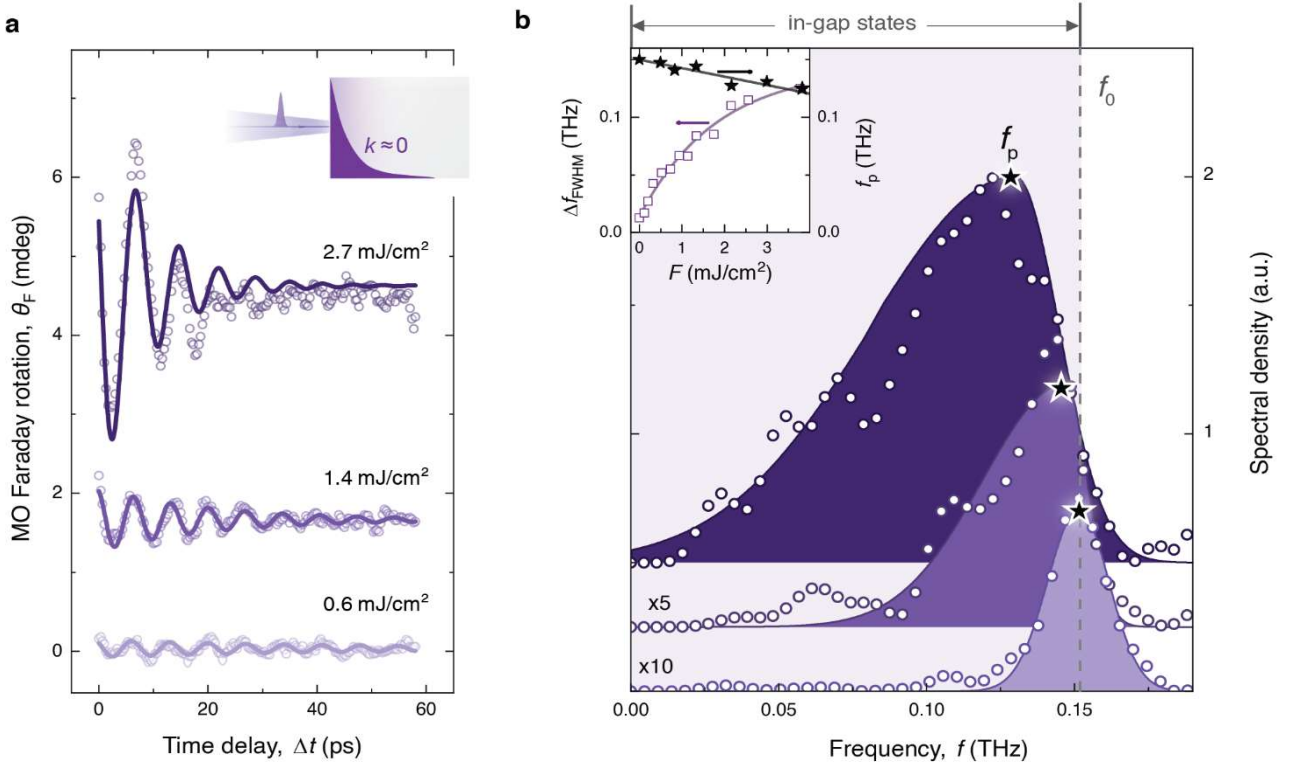

**Fig. 2 | Photoinduced continuum of in-gap magnon states. (a)** Time-resolved traces of the magneto-optical (MO) Faraday rotation $\theta_F$ showing the coherent magnon dynamics at $k \simeq 0$ for low ($F = 0.6$ mJ/cm$^2$), moderate ($F = 1.4$ mJ/cm$^2$) and strong ($F = 2.7$ mJ/cm$^2$) fluences of the above-bandgap pump excitation ($h\nu = 3.1$ eV). Circles are experimental data points; solid lines are the best fits with damped sine function. The unmatching between experimental data points and fits under high fluence indicates the asymmetrical reshaping of the spectra of the signal. The inset schematically shows spatial distribution of the measured dynamics localized within the optical penetration depth $\delta$. **(b)** FFT transform showing the spectral density distribution of the time-traces from (a). Stars mark the peak frequencies $f_p$. Grey dashed line indicates magnon gap $f_0$, light-purple colour-filled area below $f_0$ represents in-gap states. Inset shows the evolution of the bandwidth $\Delta f_{FWHM}$ (purple squares, left axis) and peak frequency $f_p$ (black stars, right axis) of the excited magnon dynamics as a function of pump fluence.

**Fingerprints of a photoengineered magnon band**

Next, we investigate how the *CT*-excitations affect magnon components with a finite momentum $k > 0$. These non-uniform components arise from the strong sub-wavelength localization of the *CT*-excitation and are characterized by a finite bandwidth $\Delta k \sim \delta^{-1}$.[9,42,43] The corresponding finite-$k$ excitations have non-zero group velocities $V_k = \partial f / \partial k \neq 0$ and thus rapidly escape the photoexcited region. Nonetheless, their early-time dynamics within $\tau_d \sim \delta / V_k$ may still provide valuable insights into the finite-$k$ components of the magnon spectrum under the electronic excitation. To capture their dynamics, we performed time-resolved measurements of the magneto-optical Kerr effect (MOKE) in reflection geometry (Extended Data Fig. 1). In this configuration, MOKE provides access to a finite out-of-plane magnon momentum $k_0$, which at normal incidence is given by $k_0 = 2k^*_{pr}$, where $k^*_{pr}$ is the wavevector of the probe ray inside the material[9,45].

Figure 3a shows a time-resolved trace of the MOKE rotation $\theta_K$ with an optical probe at $\lambda_{pr}$ = 515 nm, revealing the magnon dynamics at $k_0 \approx 0.93 \times 10^5$ cm$^{-1}$ ($\lambda_0 \approx 110$ nm). To compare the spectral response of $k = k_0$ magnons with those of $k \simeq 0$, we mapped their Fourier spectra – acquired at the same temperature and pump fluence – onto the equilibrium magnon dispersion curve, $f(k)$, see Fig. 3b. While the finite-$k$ spectrum is dominated by the magnon component at the frequency $f(k_0) = f_k \approx 0.25$ THz, a broad spectral response centred around 0.1 THz is also clearly observed. This low-frequency response, substantially lower than both $f_k$ and $f_0$, closely matches the central frequency $f_p$ of the spectrally renormalized oscillations at $k \simeq 0$. The presence of spectral weight centred at $f_p < f_0$ at both $k \simeq 0$ and $k = k_0$ suggests the formation of a strongly non-equilibrium magnon band around $f_p$. This band splits off from the original magnon spectrum and represents an electronically modified magnetic state. Being localized within the pump penetration depth $\delta$, this band extends significantly to higher momenta $k$.

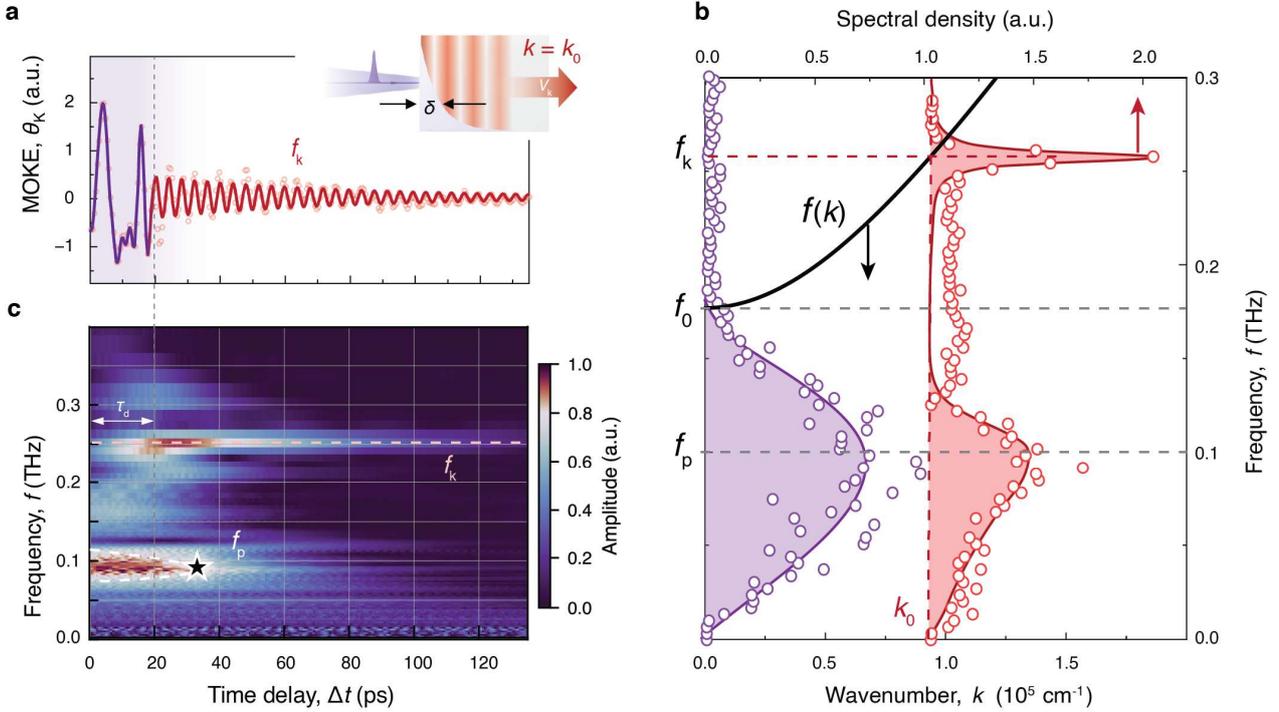

**Fig. 3 | Fingerprints of a photoengineered magnonic band. (a)** Time-resolved trace of the magneto-optical Kerr (MOKE) rotation $\theta_K$ probing the coherent dynamics of finite-$k$ components of the magnon packet excited by the above-bandgap pump excitation ($h\nu$ = 3.1 eV). The probe wavelength set to 515 nm ensures selective detection of components at $k = k_0 = 0.93 \times 10^5$ cm$^{-1}$ with the frequency $f_k \simeq 0.25$ THz. The shaded purple area highlights a distortion in the early-time dynamics of magnons at $k = k_0$. The gray dashed vertical line shows the early time $\tau_d$ in the $f_k$ mode. Top-right inset schematically shows the spatial distribution of the probed magnon dynamics; $\delta$ is the pump penetration depth; $V_k$ is the group velocity of the propagating magnon. **(b)** Fourier spectrum of simultaneously measured $k \simeq 0$ mode (purple, MO Faraday rotation) and $k = k_0$ (red, MO Kerr rotation) mapped on top of the equilibrium dispersion curve $f(k)$ (black solid line). Red dashed line marks $f_k$. Both of red and purple colour-filled areas below $f_0$ represents in-gap states. Upper (lower) grey dashed line marks the magnon gap $f_0$ (peak frequencies $f_p$). **(c)** Wavelet analysis of the signal from (a). Pink dashed line indicates the frequency $f_k$. White dashed line highlights the evolution of the spectral density characterized by the peak frequency $f_p$ in the early-time dynamics. White arrow shows the early time $\tau_d$ in the $f_k$ mode. The experiments were performed at $T$=65 K.

To understand the relationship between the spectral components at $f_p$ and $f_k$, we performed a wavelet analysis of the finite-$k$ magnon dynamics, see Fig. 3c. This analysis reveals that while the spatially localized signal at $f_p$ emerges immediately after excitation, the spectral response at $f_k$, corresponding to propagating magnons, is delayed by $\tau_d \approx 20$ ps. This delay also manifests as an early-time distortion in the time-resolved waveforms (highlighted in Fig. 3a) and shows a strong correlation with the pump's penetration depth, $\delta$ (see Extended Data Fig. 3). The delay suggests that the finite-$k$ magnons originate from the electronically excited region, and being emitted freely propagate into the bulk of the sample without any spectral modification. Indeed, our experiments show that – unlike the behaviour of localized components at $f_p$ – the frequency and the decay rate of the $f_k$ propagating component are independent on the pump fluence, see Extended Data Fig. 4. In this scenario, $\tau_d$ represents the time needed the finite-$k$ magnon to traverse the excited region, $\delta$, and thus can be used to estimate the magnon velocity $V_k^*$ within this electronically excited region. Surprisingly, our estimates show that $V_k^* \simeq \delta/\tau_d \approx 5$ nm/ps – roughly half of the nominal $V_k \approx 12$ nm/ps for free-space magnon propagation, suggesting that the optically excited region exhibits reduced group velocity and thus the exchange interaction.

**Localized state with renormalized magnetic interactions**

To establish *CT*-driven renormalization of the exchange interaction as the primary mechanism behind the emergence of the photoinduced magnon band, we consider a minimal model that replicates the observed magnon behaviour in DyFeO$_3$. The model is a 3D bipartite spin lattice divided into sublattices 1 and 2, with principal magnetic interactions described by the microscopic Hamiltonian[37]:

$$H = J_0 \sum_{\substack{i \in 1 \\ \delta \in \{\pm \hat{x}, \pm \hat{y}, \pm \hat{z}\}}} \vec{S}_i \cdot \vec{S}_{i+\delta} + D_0 \sum_{\substack{i \in 1 \\ \delta \in \{\pm \hat{y}\}}} (S^z{}_i \cdot S^x{}_{i+\delta} - S^x{}_i \cdot S^z{}_{i+\delta}) + \sum_{i \in 1,2} \left[ K_2 (S^y{}_i)^2 - \frac{K_4}{S^2} (S^y{}_i)^4 \right] \quad (2)$$

The parameters $J$ and $D$ define the strength of the AFM superexchange and DM couplings, respectively. $K_2$ and $K_4$ represent the strengths of quadratic, and respectively quartic, single-ion magnetic anisotropies defining the orientation of the iron spins. For exact values of the parameters, details of the considered Hamiltonian and the extended discussion of magnetic properties in thermal equilibrium see Supplementary Note S1.

To simulate the effect of the *CT*-excitation on the magnon spectrum, we introduce an effective spatio-temporal quench $Q(z,t)$ of the equilibrium exchange interaction $J_0$, expressed as $J(z,t) = J_0 (1 - Q(z,t))$. To estimate the temporal characteristics of the quench, we measured the dynamics of transient reflectivity, which is sensitive to the electronic subsystem. Our experiments indicate that the lifetime of the excitations extends well beyond the period of spin precession, see Extended Data Fig. 6. To align with these observations, we take the quench profile $Q(z,t)$ to be a step function in time and exponentially decaying in space, along the sample normal, see Fig. 4a. We also assume that the quench magnitude $Q_0$ scales linearly with the pump radiation intensity and, consequently, with the density of photoexcited carriers. This model provides insights into the coherent dynamics both within the excitation region and outside of it, where the sample remains unaffected but can still host magnons propagating outward, as shown in Fig. 4b.

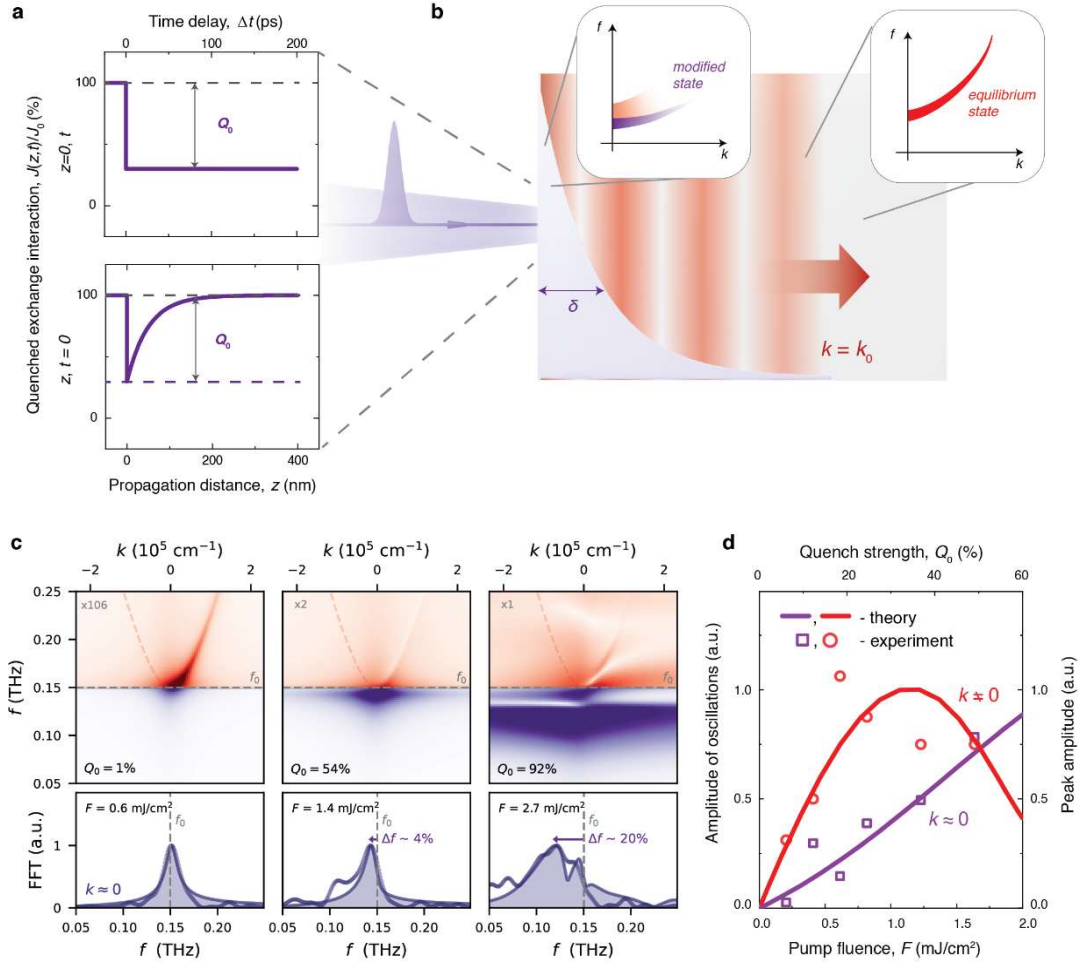

**Fig. 4 | Ultrafast quench of exchange interaction. (a)** (top panel) Temporal and (bottom panel) spatial profile of quenching of the exchange interaction $J(z,t)/J_0$. $Q_0$ indicates the maximum quenching strength reached at the sample's surface. **(b)** The schematic concept of the model considering that ultrafast *CT*-excitation, via quenching of $J$, leads to the forming of the modified magnetic state which is localized in the excited area. Nevertheless, in the bulk, where $J$ is unaffected, the magnetic state is equilibrium. Two insets schematically show the dispersions of modified and equilibrium states. **(c)** (top panel) Reciprocal space snapshots of the numerically calculated coherent spin dynamics in the arbitrary scale (magnification is specified in the top-left corner) for different quenching strength $Q_0$. Red dashed line shows the equilibrium dispersion of spin waves. The straight grey dashed line shows the equilibrium magnon gap $f_0 = 0.15$ THz. (bottom panel) Corresponding linear cuts (solid lines with filled area) from the snapshots at $k = 0$ from the top panel mapped on top of the experimental Fourier spectra (circles) of the $k \approx 0$ dynamics presented in Fig. 2b. The straight grey dashed line shows the equilibrium magnon gap $f_0 = 0.15$ THz. The purple arrow highlights the excitation-induced red shift of the spectral peak $\Delta f$ which is observed in both of experiment and theory. The exact values of $Q_0$ were chosen for the best fit of the experimental data by the numerical calculations. **(d)** Experimental values of the amplitude of oscillations as a function of pump fluence (symbols; left-bottom axes) plotted on top of the theoretical amplitude of the spectral peak as a function of the quench strength $Q_0$ (lines; right-top axes) for both of $k \approx 0$ (purple) and $k \neq 0$ (red) magnon states.

The results of modelling spin dynamics for various values of $Q_0$ are shown in Fig. 4c, top panel. Here, we present the quench-induced spatio-temporal profiles of coherent magnon dynamics as a two-dimensional (2D) Fourier transform, mapped onto frequency and momentum coordinates. Such representation not only provides access to the spectral distribution of the excited magnons but also enables a straightforward mapping onto the magnon dispersion $f(k)$, see Eq. (1). Our modelling shows that suppressing of $J$ results in the off-surface emission of coherent magnons, with $k$-space distribution defined by the spatial profile of the quench. In the weak-pumping regime (Fig. 4c, left panel), the main effect is a small perturbation of the spins near the sample surface. As a result, the quench induces magnon emission off the surface, with spectral features strictly following the equilibrium dispersion relation $f(k)$. In contrast, in the stronger-pumping regime (Fig. 4c, middle and right panels), a broad spectral band clearly emerges. This band extends significantly in $k$-space and shifts to frequencies well below $f_0$, resembling the experimental results from Fig. 3c. To estimate the magnitude of the exchange quench reached in our experiment, we compare the theoretical modelling with our fluence-dependent measurements. Specifically, we take the line cuts along $k = 0$ from the calculated spectra and map them to the measured spectra from Fig. 2b. The comparison (Fig. 4, bottom panel) demonstrates that the theoretical fits accurately reproduce key spectral features of the experimental observations: the asymmetric broadening and the redshift of the peak below $f_0$. Surprisingly, in the strong-pumping regime ($F \simeq 2.7$ mJ/cm$^2$) the best agreement between theory and experiment is achieved by assuming that the exchange interaction at the material's surface is quenched by about 90%. Indeed, at the applied fluence, the photoexcitation generates approximately 10% of charge carriers per unit cell – enough to almost fully suppress the exchange interaction, see Supplementary Note S2.1.

Our theoretical model further supports the proposed large-magnitude exchange quenching scenario, accurately capturing the dependence of both $k \simeq 0$ and finite-$k$ magnon amplitudes on pump fluence, see Fig. 4d. Notably, both the model and experiment reveal unexpected deviations from a simple linear relationship: while the $k \simeq 0$ mode exhibits super-linear growth, the amplitude of the propagating $k \neq 0$ magnon mode saturates and eventually undergoes suppression. In particular, our analysis shows that the amplitude suppression results from the local trapping of spin-wave excitation within the electronically excited near-surface region, where the magnetic potential undergoes significant modification, for details see Supplementary Note S3.2.

To finally confirm that exchange quenching is the key factor in photoengineering the spin-wave spectrum, we also theoretically explore the ultrafast quench of other magnetic interactions. First, our simulations show that quenching the DM interaction $D$ has a similar effect on the magnon spectrum as the quenching of $J$, see Supplementary Note S3.3. This may be a consequence of the coupling of the weak ferromagnetic moment **M** to the Néel order via DM interaction (the second term in Eq. (2)). However, quenching $D$ alone fails to fully reproduce both the spectral lineshape and the fluence-dependent amplitude behaviour observed in the experiment. Second, the quench of the anisotropies – $K_2$ and $K_4$ – does not even initiate the coherent dynamics of spins, see Supplementary Notes S3.1 and S3.4. Lastly, we consider another effect accompanying *CT*-excitation – charge carrier delocalization – which could influence the spin-wave spectrum by suppressing the magnitude of local magnetic moments $S$. Our calculations show that although the reduction of $S$ impacts the magnon spectra, it produces line shapes entirely different from those observed in our experiments, see Supplementary Note S3.5.

**Conclusions and Outlook**

Our findings extend the scope of resonant optical control in AFMs by demonstrating that resonant pumping of *CT* electronic transitions – strongly coupled to fundamental spin exchange – can effectively reshape the magnon spectrum. Our results show that, rather than being limited to exploiting nonlinear magnon-magnon interactions[11–14,45,46], the renormalization of the magnon spectrum can be achieved by directly targeting electronic resonances without destroying long-range magnetic order. Since the majority of insulating antiferromagnets belong to the class of *CT* systems, our discovery can be applied to a wide range of AFM materials[21,47] and may also provide insight into the microscopic mechanisms behind the optical generation of broadband THz magnon wavepackets in various iron oxides[9,42,43]. Nanoscale control of fundamental magnetic interaction, including exchange, opens new avenues for THz magnonics and spintronics, enabling the long-sought creation of dynamic AFM magnonic crystals[48,49]. On a fundamental level, our approach may enable enhanced dynamic interactions, such as magnon-magnon and magnon-phonon couplings, which are typically constrained by energy conservation limitations[50].

**Materials and Methods**

**Sample and experimental set-up**. We studied a monocrystalline, 60-μm-thick $DyFeO_3$ sample grown by floating-zone melting. The sample was cut perpendicularly to the *z* crystallographic axis in the form of a thin slab. The sample was kept in a closed-cycle He cryostat (Montana Instruments), which allowed it to cool down to 10 K and vary the temperature with high stability in a wide temperature range (10–250 K). An amplified 1 kHz Ti:Sapphire laser system (SpitfirePro, Spectra Physics; central wavelength, 800 nm; pulse energy, 4 mJ; pulse duration, 50 fs) forms the basis of the experimental setup. A large fraction of this output is used to pump an optical parametric amplifier (TOPAS TWINS, LIGHT CONVERSION) and thus to produce pulses at variable photon energy. A small portion of the pulses was sent through a mechanical delay line and used as a probe of the spin dynamics in the reflection and transmission geometries. Pump and probe pulses were focused onto the $DyFeO_3$ sample (pump spot diameter, 300 μm; typical fluence, 1 mJ/cm$^2$; probe spot diameter, 80 μm). The pump-induced changes in the polarization $\theta_{K,F}$ of the reflected or transmitted probe pulse were measured using an optical polarization bridge (Wollaston prism) and a pair of balanced Si photodetectors.

**Numerical modelling**. The system of coupled evolution equations (for the derivation see Supplementary Note S2.2) was numerically solved for every site in a system of size N = 5000 unit cells (corresponding to 3.8 μm), using the Euler cluster of ETH Zürich. The initial conditions for the equations were taken to be those of the equilibrium canted phase. At the sample edge facing the pump, boundary conditions (one fewer neighbour for each surface site, compared to the bulk) introduce small variations of the equilibrium canting angle, which manifest as a domain wall of approximate size 5 unit cells. We determine this configuration numerically, by minimizing the energy in Dirac's Lagrangian method, prior to applying the parameter quench and starting the real-time evolution.

Since the experimental sample is 60 μm thick and the focus of this work is on near-surface effects arising from strong absorption of the pump radiation, we can effectively treat the material as semi-infinite in the $\hat{z}$ direction. At the far edge of the simulated system, we therefore employ an absorbing boundary condition, to avoid any unphysical reflection of the propagating wavepackets. This is implemented by adding $N_{\text{boundary}} = 250$ extra sites with progressively higher values of the local damping $\lambda$. Afterwards, these sites are not taken into account when recovering the nonequilibrium response of the system.

**Acknowledgements**

We thank K. Saeedi Ilkhchy and C. Berkhout for their technical support, and J. Mentink for fruitful discussions. This work is supported by ERC grant 852050 MAGSHAKE; ERC grant 101078206 ASTRAL; programme "Materials for the Quantum Age" (QuMat, registration number 024.005.006), which is part of the Gravitation programme financed by the Dutch Ministry of Education, Culture and Science (OCW); and the European Research Council ERC grant agreement no. 101054664 273 (SPARTACUS). R.C. received support by Horizon Europe EIC Pathfinder under the grant IQARO number 101115190 and from the Italian PNRR MUR project PE0000023-NQSTI. R.A. and E.D. acknowledge the SNSF project 200021_212899, NCCR SPIN, a National Centre of Competence in Research, funded by the Swiss National Science Foundation (grant number 225153), and the Swiss State Secretariat for Education, Research and Innovation contract number UeM019-1). The work was partly was supported by the Swiss State Secretariat for Education, Research and Innovation (SERI) under contract no. MB22.00071, the Gordon and Betty Moore Foundation (grant no. 332 GBMF10451 to A.D.C.), the European Research Council (ERC) (grant no. 677458), by the project Quantox of QuantERA ERA-NET Cofund in Quantum Technologies, and by the Netherlands Organisation for Scientific Research (NWO/OCW) as part of the VIDI (project 016.Vidi.189.061 to A.D.C.), the ENW-GROOT (project TOPCORE) program (to A.D.C.)

**Contributions**

D.A. and A.D.C. conceived the project. D.A., A.V.K. and A.D.C. supervised the study. V.R., J.R.H., M.X.N. and D.A. performed the experiments. R.A. and E.D. provided theoretical model. B.A.I, R.V.M., S.C. and R.C. contributed to the theoretical treatment of the experimental results. D.A., V.R. and R.A. wrote original draft with feedback from all the co-authors.

**Competing interests**

The authors declare no competing interests.

**Data availability**

Source data for figures are publicly available at [Zenodo](). All other data that support the findings of this paper are available from the corresponding authors upon request.

**Code availability**

The code used to simulate the magnetic dynamics is available upon reasonable request.

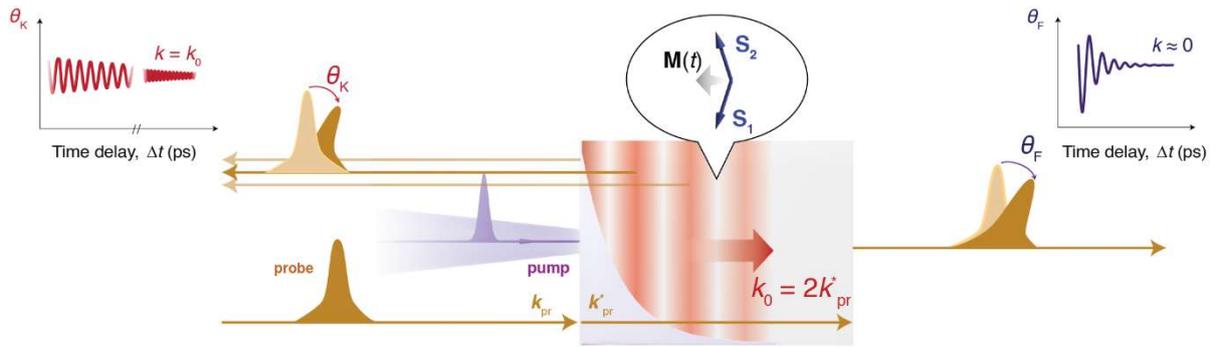

**Extended Data Fig. 1 | Schematics of the experiment.** The time-resolved signals (illustrated schematically at the top-left and top-right) were obtained using a pump-probe technique. The first, pump pulse (purple) excites the near-surface region (depicted as the blue area with exponential profile), and the second, probe pulse with wavevector $k_{pr}$ (orange, bottom-left) measures the magnetic response. When transmitted through the medium, the probe pulse experiences a magneto-optical Faraday rotation $\theta_F$ of its polarization due to the net magnetization $\mathbf{M}(t)$ (shown enlarged in the inset), thus providing information about localized magnon dynamics with $k \approx 0$ (middle-right; purple colour). The refracted portion of the probe pulse acquires a new wavevector $k^*_{pr}$ within the sample, and is subsequently reflected by inhomogeneities possessing a spatial periodicity matching – specifically, magnons with the wavevector $k_0 = 2k^*_{pr}$. The polarization rotation of this reflected portion is the magneto-optical Kerr rotation $\theta_K$, which provides insights into propagating magnons with wavevector $k_0$ (top-left; red colour).

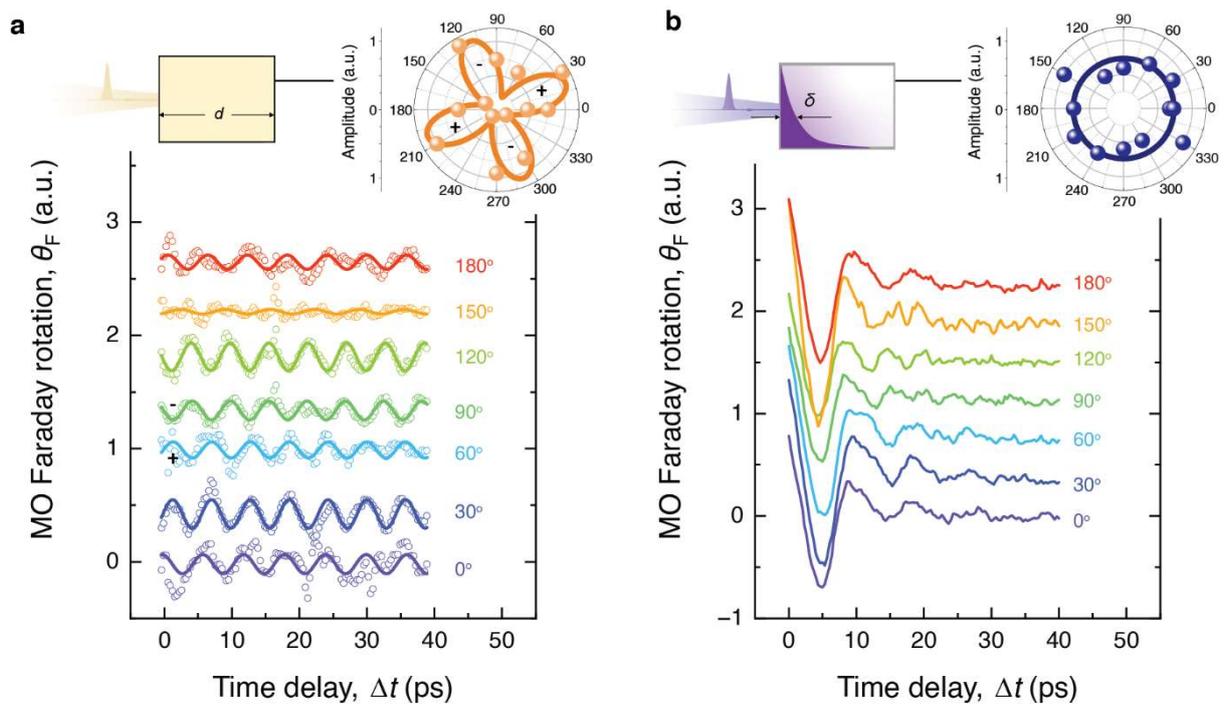

**Extended Data Fig. 2 | Pump-polarization dependence.** Time-traces for the $k \approx 0$ dynamics excited by linearly-polarized pump with different orientation of polarization when pumping **(a)** not in *CT*-regime and **(b)** in *CT*-regime. On the panel (a) the scatters are experimental data points and solid lines are sinusoidal fits. On panel (b) solid lines are the experimental time-traces. Labels from $0^0$ to $180^0$ denote the orientation of the pump polarization for every particular curve. Top-left insets on both graphs show the spatial distribution of the excitation. Top-right insets are the dependences of the amplitude from the angle of the pump polarization. While in *CT*-regime (b) the magnetic dynamics is isotropic, in off-resonant regime (a) the orientation of pump polarization allows to control not only the amplitude of the dynamics, but also the phase (compare "+" light-blue and "-" dark-green curves).

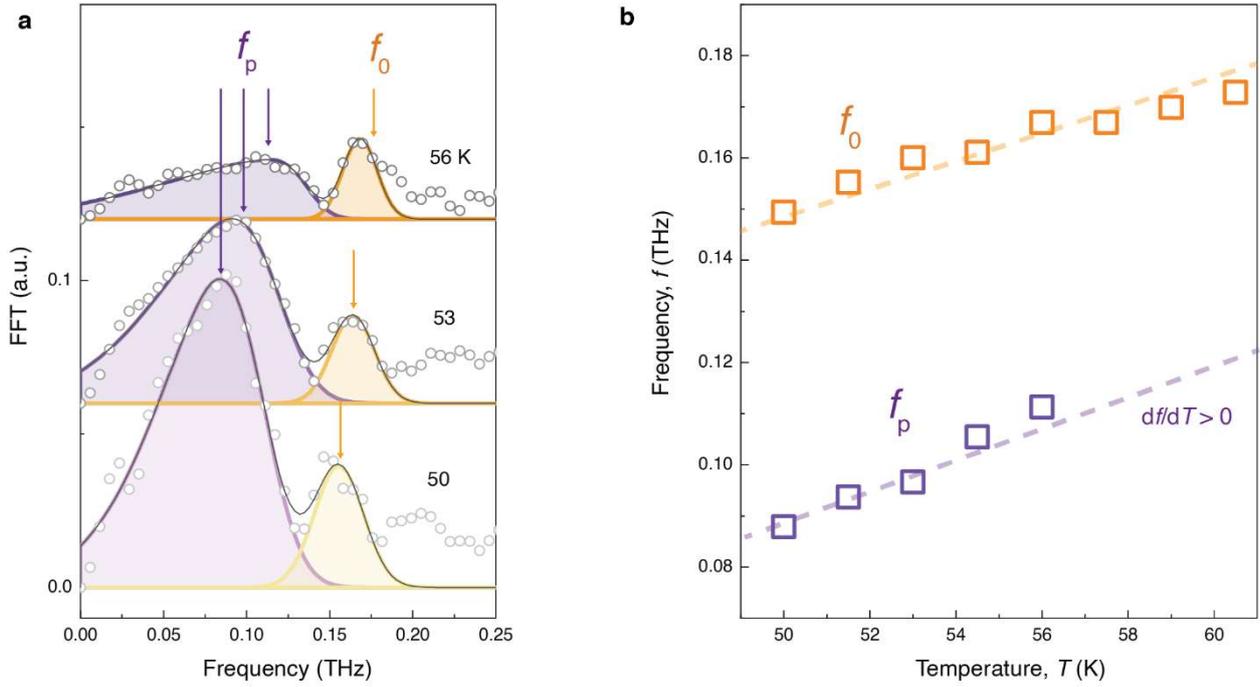

**Extended Data Fig. 3 | Temperature-evolution of the localized $k \approx 0$ magnon state. (a)** Fourier spectra of the unperturbed equilibrium (orange colour) and *CT*-excitation-modulated part (purple) part of the localized $k \approx 0$ magnon mode for different temperatures ($T = 50, 53, 56$ K). Black solid lines are the fits taken as a sum of asymmetrical gaussian (purple filled peaks) and symmetrical gaussian (orange filled peaks) at frequencies $f_P$ and $f_0$, respectively. **(b)** The frequencies $f_0$ and $f_P$ plotted as a function of temperature. Squares are experimental values, dashed lines with positive slope ($df/dT > 0$) are guide-to-eye.

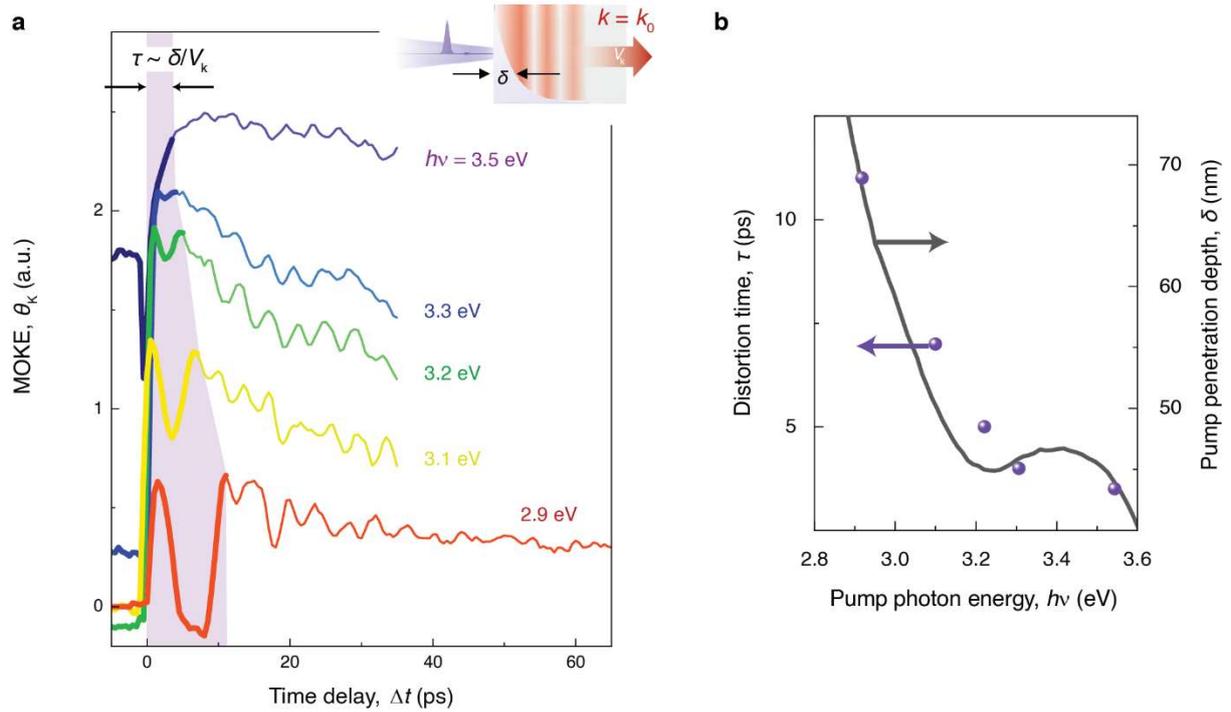

**Extended Data Fig. 4 | The effect of pump photon energy on the duration of early-time distortion in the dynamics of propagating finite-*k* components. (a)** Time-resolved traces of the magneto-optical Kerr rotation (MOKE) $\theta_k$ excited by pumping with different pump photon energies $h\nu$ from 2.9 eV up to 3.5 eV. The bold part of the curves, as well as purple shaded area, highlights the distorted early-time magnon dynamics with a duration of $\tau$. Top right inset shows a spatial distribution of the probed magnon dynamics; $\delta$ is a pump penetration depth. **(b)** $\tau$ (left axis, circles) and $\delta$ (right axis, solid line) as a function of pump photon energy $h\nu$. Here $\tau$ was determined by visual inspection of the start-time of coherent oscillations from (a).

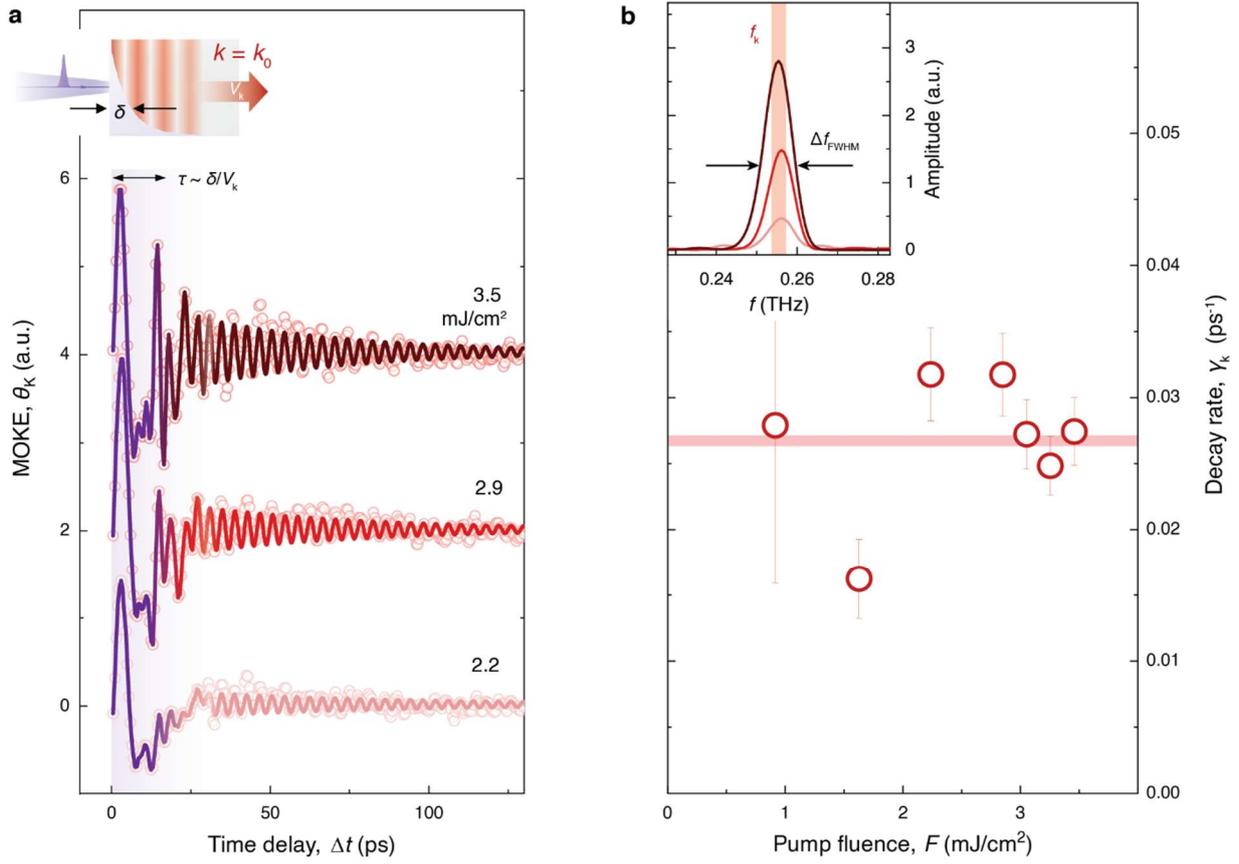

**Extended Data Fig. 5 | Pump fluence-independent decay rate of propagating finite-$k$ magnons.** (a) Time-traces of $k = k_0$ magnon mode for different fluences of the above-bandgap pump excitation ($h\nu$ = 3.1 eV). The inset schematically shows spatial localization of the measured dynamics; $\delta$ is the pump penetration depth; $V_k$ is the group velocity of the propagating magnon. Purple shady area, as well as purple colour of curves, highlights early-time $\tau$ distorted dynamics. (b) (circles) Experimental values of the decay rate of *CT*-driven $k = k_0$ magnon mode, $\gamma_k$, as a function of pump fluence. Red solid line is guide-to-eye. (inset) Fourier peak of late-time coherent part of $k = k_0$ mode for different pump fluences. Neither the central frequency of the peak $f_k \sim 0.25$ THz (highlighted by shady red stripe), nor the FWHM bandwidth $\Delta f_{FWHM} \sim 0.01$ THz is unaffected by the pump-fluence.

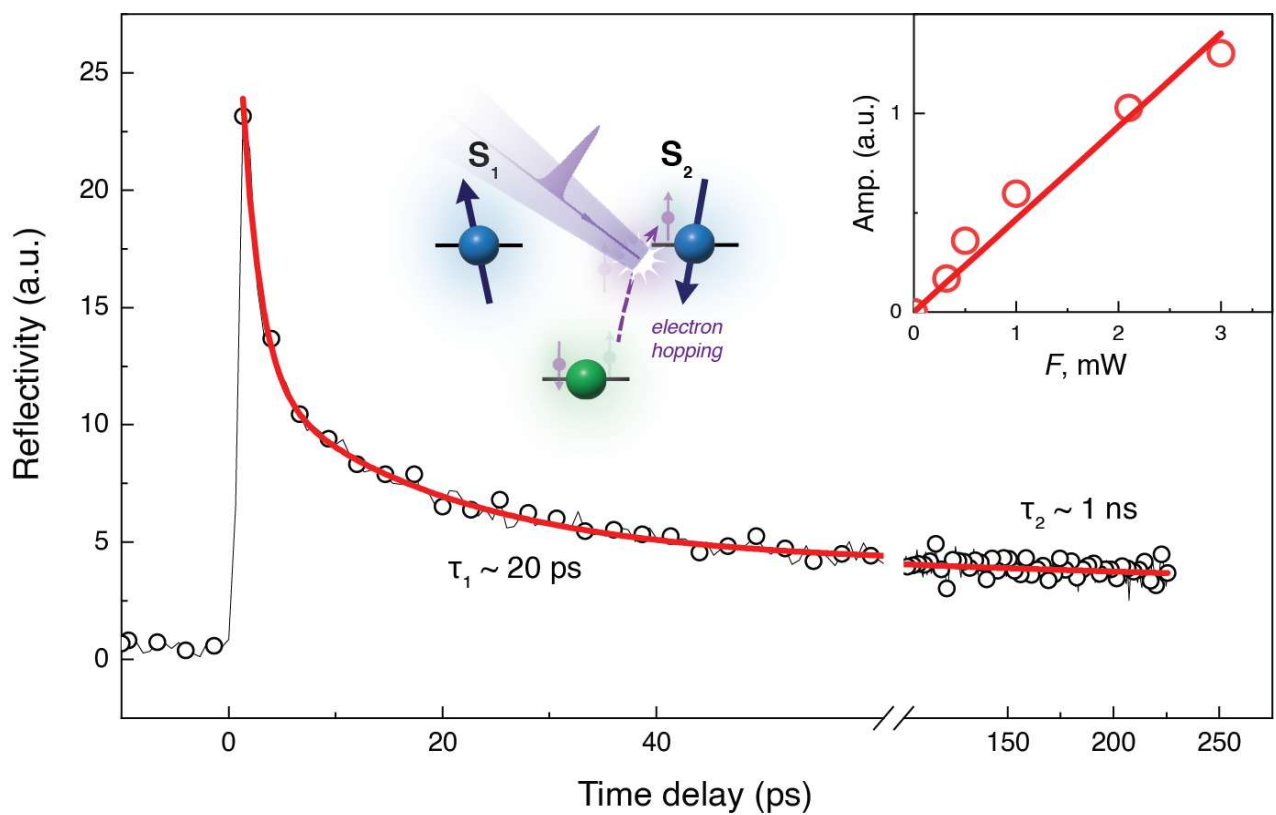

**Extended Data Fig. 6 | Lifetime of *CT* electronic excitation induced by light.** Time-resolved dynamics of the transient reflectivity following the ultrafast *CT* excitation. Circles represent experimental data points, and the red line shows a three-exponential fit with relaxation times: $\tau_0 \sim 1$ ps, $\tau_1 \sim 20$ ps and $\tau_2 \sim 1$ ns. The last two times indicate that the non-equilibrium behaviour of the photoexcited carriers last significantly longer than the lifetime of the spin precession (~ 7 ps). The central inset is a schematic representation of the *CT* electronic excitation. The top-left inset is the fluence-dependence of the amplitude of the $\tau_1$ exponent. Open circles are the experimental values, and straight solid line is a linear fit.

Supplementary information

# Photoengineering the Magnon Spectrum in an Insulating Antiferromagnet

V. Radovskaia[1]*, R. Andrei[2], J.R. Hortensius[3,4], R.V. Mikhaylovskiy[5], R. Citro[6], S. Chattopadhyay[2,7], M.X. Na[1], B.A. Ivanov[1,8], E. Demler[2], A.V. Kimel[1], A.D. Caviglia[9], and D. Afanasiev[1]*

[1] *Radboud University, Institute for Molecules and Materials, 6525 AJ Nijmegen, Netherlands*

[2] *Institute for Theoretical Physics, ETH Zurich, 8093 Zurich, Switzerland*

[3] *Kavli Institute of Nanoscience, Delft University of Technology, P.O. Box 5046, 2600 GA Delft, Netherlands*

[4] *Electromagnetic Signatures and Propagation, TNO, 2597 AK The Hague, The Netherlands*

[5] *Department of Physics, Lancaster University, Bailrigg, Lancaster LA1 4YW, United Kingdom*

[6] *Dipartimento di Fisica "E. R. Caianiello", Università degli Studi di Salerno and CNR-SPIN, Via Giovanni Paolo II, I-84084 Fisciano (Sa), Italy*

[7] *Lyman Laboratory, Department of Physics, Harvard University, Cambridge, MA 02138, USA*

[8] *Institute of Magnetism, National Academy of Sciences of Ukraine, Kiev, 03142 Ukraine*

[9] *Department of Quantum Matter Physics, University of Geneva, CH-1211 Geneva 4, Switzerland*

*Correspondence to: viktoriia.radovskaia@ru.nl, d.afanasiev@science.ru.nl

# Table of Contents

# S1. Static and dynamic magnetic properties of DyFeO3 in equilibrium

## S1.1. Microscopic magnetic Hamiltonian

To understand the mechanism behind the appearance and driving of the nonequilibrium in-gap excitations, we start from a minimal spin model that reproduces the equilibrium physics.

Since previous studies[1,2] have emphasized the crucial role of anisotropy and strong spin-orbit interactions for the description of the possible magnetic orderings in DyFeO3, in our model we consider the following contributions: nearest neighbour AFM exchange $J$, a Dzyaloshinskii-Moriya (DM) interaction of strength $D$ acting along $\hat{y}$ bonds, and quadratic ($K_2$) as well as quartic ($K_4$) single-ion $\hat{y}$−axis anisotropies. Working with a 3D square lattice of spins $S = 5/2$, which is split into sublattices **1** and **2**, the Hamiltonian is:

$$H = J \sum_{\substack{i \in \mathbf{1} \\ \delta \in \{\pm\hat{x},\pm\hat{y},\pm\hat{z}\}}} \vec{S}_i \cdot \vec{S}_{i+\delta} + D \sum_{\substack{i \in \mathbf{1} \\ \delta \in \{\pm\hat{y}\}}} (S_i^z \cdot S_{i+\delta}^x - S_i^x \cdot S_{i+\delta}^z) +$$

$$+ \sum_{i \in \mathbf{1,2}} \left[ K_2 (S_i^y)^2 - \frac{K_4}{S^2} (S_i^y)^4 \right] - \boldsymbol{H_0} \cdot \sum_{i \in \mathbf{1,2}} \vec{S}_i \qquad (1)$$

where $\boldsymbol{H_0}$ represents an external magnetic field.

Previous treatments[1,2] rely on coarse-graining the magnetic order parameters over length scales large compared to the lattice constant. However, the focus of this work is the strongly-localized spectrum renormalization that pumping induces at the sample surface. Significant spatial variation of the pump's intensity, over relatively few lattice constants near the edge, motivates the employment of a microscopic Hamiltonian description instead. The connection between the microscopic and coarse-grained approaches can be made as follows: assume slow variations of the spin orientations for each sublattice separately, and define the locally averaged order parameters:

$$\boldsymbol{m} = \langle \vec{S}_1 \rangle + \langle \vec{S}_2 \rangle \quad \text{and} \quad \boldsymbol{l} = \langle \vec{S}_1 \rangle - \langle \vec{S}_2 \rangle \qquad (2)$$

where $\boldsymbol{m}$ is the net magnetization given by the sum of the magnetizations of the antiferromagnetically coupled magnetic sublattices, and $\boldsymbol{l}$ is the antiferromagnetic Néel vector.

For a system consisting of $N_0$ unit cells (i.e. $2N_0$ sites in total), with spatially uniform order and assuming $m_y = 0$, the energy density expectation is given by

$$\frac{\langle H \rangle}{N_0} = 3J\boldsymbol{m}^2 + D(l_z m_x - l_x m_z) + \frac{K_2}{2} l_y^2 - \frac{K_4}{8S^2} l_y^4 - \boldsymbol{H_0} \cdot \mathbf{m} - 6JS^2 \qquad (3)$$

which has the same form as energy densities used in refs.[1,2].

In the following, we will work in the microscopic Hamiltonian picture, and consider the case without external magnetic field $\boldsymbol{H_0} = 0$. Starting with the Eq. (1), we use Schwinger bosons to represent states of the spin−5/2 system, and the Landau-Lifshitz approximation to dynamics, which is equivalent to assuming that the many-body spin state factorizes between different sites:

$$|\psi\rangle = \prod_j \prod_{\alpha=1}^{5} \left[ \cos\left(\frac{\pi}{4} - \frac{\theta_j}{2}\right) |\uparrow\rangle_{j,\alpha} + (-1)^j e^{i\varphi_j} \sin\left(\frac{\pi}{4} - \frac{\theta_j}{2}\right) |\downarrow\rangle_{j,\alpha} \right] \qquad (4)$$

In the above, the index $j$ runs over all lattice sites, while $\theta_j$ and $\varphi_j$ are interpreted as canting and respectively azimuthal angles at every site. The states $|\uparrow\rangle$ and $|\downarrow\rangle$ form a basis for the spin Hilbert space corresponding to each of the five electrons contributing to the on-site $S = 5/2$, and the $\alpha$ index runs over said electronic contributions.

### *S1.2. Magnetic interactions strength*

Restricting the ansatz to configurations where all spins on a single site are aligned will effectively project the dynamics of $S = 5/2$ onto an $SU(2)$ submanifold. In equilibrium, this recovers both the collinear AFM, as well as the canted state, depending on the strength of $K_2$. Specifically, the following local minima of the energy are found:

• *Canted state*. Néel ordering along the $\hat{x}$ −axis is recovered by letting all $\varphi_j = \varphi_0 = 0$, while the canting angles take a finite value $\theta_j = \theta_0$ (see Suppl. Fig. 1a), given by

$$\theta_0 = \frac{1}{2} \arctan \frac{D}{3J} \qquad (5)$$

The energy density of this state is given by

$$\frac{\langle H \rangle_{\text{canted}}}{N_0} = -6JS^2 \sqrt{1 + \left(\frac{D}{3J}\right)^2} \qquad (6)$$

Full equations of motion, Eq. (17, 18) will be derived below. By linearizing them around the equilibrium canted state, one obtains the two magnon modes characteristic to rare-earth orthoferrites[1–3]: quasi-antiferromagnetic ($q$-AFM; see Suppl. Fig. 1b) and respectively quasi-ferromagnetic ($q$-FM; not shown here). The mechanism we propose below is only expected to excite the $q$-AFM mode, and indeed that is what we observe experimentally. Therefore, in what follows we will exclusively focus on $q$-AFM oscillations.

Anisotropy opens an energy gap $\Delta$ for excitations at $k = 0$, and so the dispersion of long-wavelength modes is described by the relation

$$\omega^2(k) = \Delta^2 + (2\pi V_m k)^2 \qquad (7)$$

where the limiting magnon velocity $V_m$ is set by $J$, $\omega(k) = 2\pi f(k)$ and $\Delta = 2\pi f_0$.

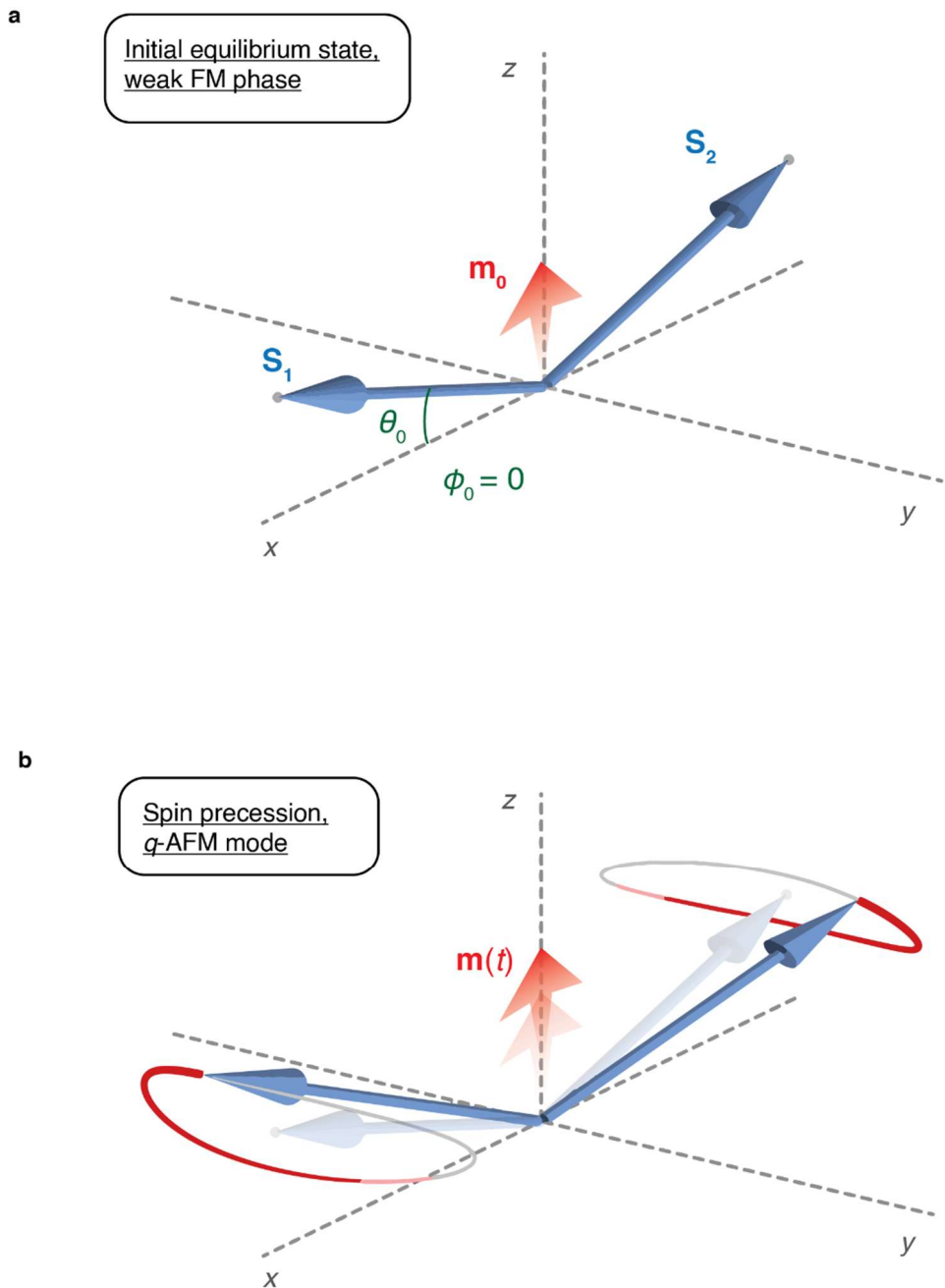

**Supplementary Figure 1: Ground state and *q*-AFM mode of the spin precession.** (a) Equilibrium state of spins $S_1$ and $S_2$ in a weak ferromagnetic (WFM) phase. $\varphi_0$ is azimuthal angle and $\theta_0$ is a canting angle which produces small net magnetisation $\mathbf{m}_0$ along the *z*-axis. (b) Schematical drawing of *q*-AFM mode leading to a time-evolution of the amplitude of magnetisation $\mathbf{m}(t)$. Ellipse-like traces show the motion of spins.

For the *q*-AFM mode in the canted phase, we obtain:

$$\Delta_{\text{canted}} = 2\sqrt{6}S\left[1+\left(\frac{D}{3J}\right)^2\right]^{\frac{1}{4}}\sqrt{J\left[K_2+3J\left(\sqrt{1+\left(\frac{D}{3J}\right)^2}-1\right)\right]} \approx 2\sqrt{6}S\sqrt{J\left(K_2+\frac{D^2}{6J}\right)}+\cdots \quad (8)$$

$$V_{\text{canted}} = 3\sqrt{2}JSa\sqrt{\sqrt{1+\left(\frac{D}{3J}\right)^2}-\frac{1}{3}\left(1+\frac{K_2+\frac{D^2}{3J}}{\sqrt{(3J)^2+D^2}}\right)} \approx 2\sqrt{3}JSa+\cdots \quad (9)$$

• *Collinear state*. Taking instead the Néel ordering along the $\hat{\mathbf{y}}$–axis, $\varphi_j = \pi/2$, and no canting $\theta_j = 0$, the energy density is

$$\frac{\langle H \rangle_{\text{collinear}}}{N_0} = -S^2[6J+2(K_4-K_2)] \quad (10)$$

while the respective gap and magnon velocity are found as

$$\Delta_{\text{collinear}} = 2S\sqrt{(6J+2K_4-K_2)(2K_4-K_2)-D^2} \quad (11)$$

$$V^0_{\text{collinear}} = 2\sqrt{3}JSa \quad (12)$$

We linearize the temperature dependence of $K_2$ around the Morin temperature,

$$K_2(T) \approx K_2(T_M)+\beta(T-T_M) \quad (13)$$

while assuming the other parameters to not depend on *T*. The first-order transition is recovered by comparing Eq. (8) with Eq. (11), and yields:

$$K_2(T_M) = K_4 - 3J\left(\sqrt{1+\left(\frac{D}{3J}\right)^2}-1\right) \quad (14)$$

which in turn gives the more familiar expression for the gap at the Morin temperature:

$$\Delta_{\text{canted}}(T_M) = 2\sqrt{6}S\left[1+\left(\frac{D}{3J}\right)^2\right]^{\frac{1}{4}}\sqrt{JK_4} \quad (15)$$

Note that the quartic anisotropy $K_4$ does not directly enter into Eq. (8) for the magnon gap in the canted phase. Indeed, this expression arises from linearizing the equations of motion around an equilibrium point with $\langle S^y \rangle = 0$, the results of which do not receive contributions from higher-order terms, such as $(S^y)^4$. Instead, the appearance of $K_4$ in Eq. (15) is only due to its role in determining the condition Eq. (14) on $K_2$ at the transition point.

Static values of all parameters in the Hamiltonian, Eq. (1), were recovered from equilibrium measurements: the limiting magnon velocity, Eq. (9, 12), gives the exchange $J$, while the canting angle, Eq. (5), sets the DM interaction strength. The magnon gap, Eq. (15), at the Morin temperature $T_M = 50$ K yields $K_4$, while $\beta$ is determined by the temperature dependence of the gap taken from ref. [2]. Numerical values are represented in Suppl. Table 1.

**Supplementary Table 1: Numerical values of the magnetic parameters included in the Hamiltonian, Eq. (1), in the equilibrium state.**

| Parameter | Equilibrium value |
|---|---|
| $J$ | 1.94 meV |
| $D$ | 0.102 meV |
| $K_4$ | 1.13 μeV |
| $K_2$ ($T_M$) | 0.243 μeV |
| $\beta$ | 0.034 μeV/K |

## S2. Model to describe magnon dynamics driven by the *CT* electronic excitation

### S2.1. Influence of photocarrier excitation on spin degrees of freedom

Considering that above-gap pumping is highly efficient at producing charge excitations, we infer that a significant density thereof will be generated near the edge of the sample, up to about 10% for the highest experimental fluences. Measurements of transient reflectivity (Extended Data Fig. 6) show that, after relaxing to a metastable state on the scale of roughly 50 ps, the system retains clearly altered optical properties for over 200 ps. Indeed, in the absence of a suitable mechanism for carrying away energy on the scale of the charge-transfer (*CT*) gap, recombination of charge excitations is expected to be strongly suppressed[4,5] and correspondingly their lifetime will be exponentially long in $E_{CT}$.

The presence of these excitations will locally modify interactions between the remaining unpaired spins, an effect which can be qualitatively understood as renormalization of the parameters in the spin Hamiltonian, Eq. (1). Originally, the effective model describing spin dynamics is obtained by starting in the electronic ground state and integrating out charge excitations. Working instead with a non-equilibrium electronic state is then expected to modify the virtual processes which give rise to spin interactions.

In principle, all four Hamiltonian parameters may simultaneously suffer changes upon pumping. However, here we will consider for clarity the effects on the magnon spectrum of separately renormalizing each of $J$, $D$, $K_2$, $K_4$. Full microscopic calculation of such renormalizations, starting from an interacting electron Hamiltonian, is left as a subject for future studies. As a possible mechanism, we note that delocalization of mobile charge excitations across the lattice generally frustrates AFM ordering. This kinetic effect has been widely studied in a variety of contexts for the Hubbard model[6–9], generally leading to the conclusion that $J$ will be reduced, due to a small density $x$ of doublons and holes, to a new value of the form $J_r \approx J - cx|t|$. Here, $c$ is a constant of order unity, and $t$ is the hopping amplitude for electrons. In the case when $|t|$ is large compared to $J$, even a low quasiparticle density $x$ can result in significant reduction of the effective exchange interaction[9,10]. For example, in a single-band Hubbard model on the square lattice, one has $J = 4t^2/U$ and $c = 4$ (see ref.[7]), yielding $J = J_0 (1 - x\, U/t)$; with $U/t \sim 10$ typical for strongly interacting systems, it follows that taking $x$ around 10% already gives the same order of magnitude for the competing exchange and kinetic terms. This, in turn, can lead to a sizable effective suppression of $J$.

### S2.2. Spatial and temporal profile of the CT-driven quench of magnetic interactions

Since strong optical absorption localizes the pump's effect near the surface, we assume that the excited quasiparticle density exponentially decays with depth $z$ in the sample. Furthermore, as the pump duration is very short compared to magnon timescales, but the excitations are rather long-lived, we make the simplifying assumption that the time dependence is a step function. This leads to the following expressions for the renormalized parameters:

$$P(z,t) = P_0(1 - Q(z,t)) = P_0 \left(1 - Q_0\, e^{-\frac{z}{\delta}} \Theta(t)\right) \qquad (16)$$

where $P = J, D, K_2$ or $K_4$, and $P_0$ is its corresponding equilibrium value; $Q(z, t)$ is a spatio-temporal quench; $Q_0$ is the relative strength of the quench, set by the pump fluence and, under weak pumping, understood to be proportional to the density of photoexcited carriers; $\delta$ is a localization length, related to the pump penetration depth; and the step function $\Theta(t)$ introduces the quench at time $t = 0$. Suppl. Fig. 2b,c display the spatial and temporal profiles of parameter quenching, for $t = 0$ and respectively $z = 0$.

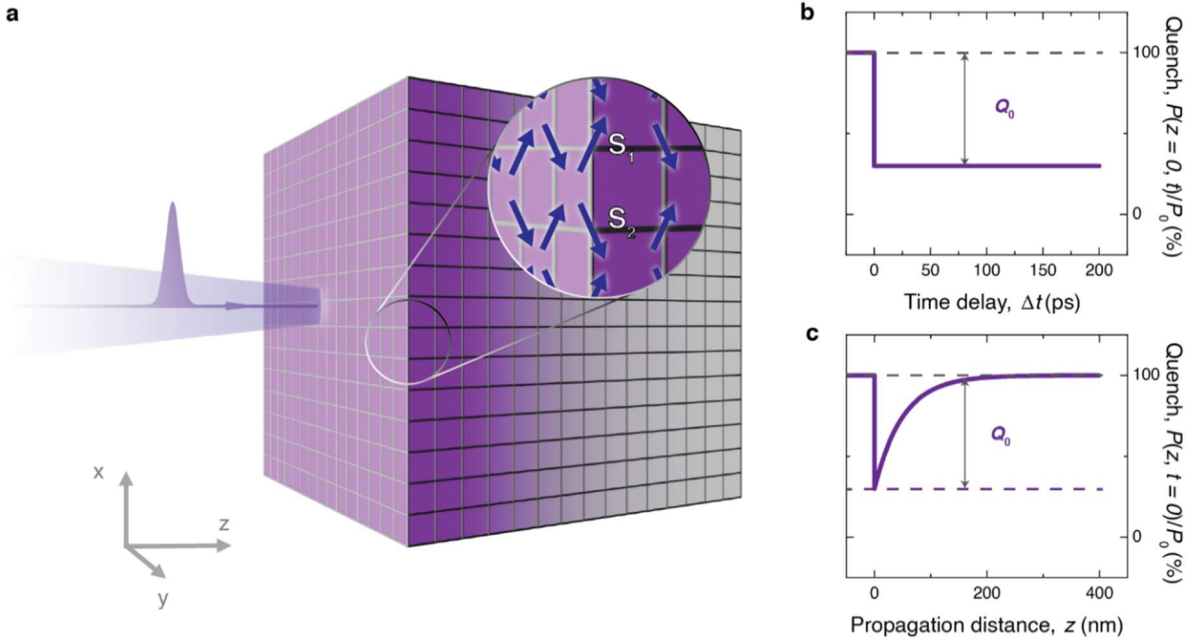

**Supplementary Figure 2: Spatial and temporal profiles of the *CT*-driven interaction quench *Q*.** **(a)** 3D square bipartite lattice of spins $S_1$ and $S_2$ used in numerical simulation. Purple color schematically highlights the profile of *CT*-excitation. **(b)** Temporal- and **(c)** spatial profile of interaction quench $P(z, t)/P_0$ with $Q_0$ indicating the maximum quenching strength reached at the sample's surface.

The post-quench dynamics is modelled using Dirac's Lagrangian method. We take a time-dependent variational wavefunction $|\psi\rangle$ as described in Eq. (4), where the angles $\theta_i$ and $\varphi_i$ start from their equilibrium values in the canted phase, and are allowed to vary with time after the quench. For large enough spot sizes of the pump, the only relevant spatial variation takes place along the $\hat{z}$ direction, so we can restrict the ansatz to be uniform in the $\hat{x} - \hat{y}$ plane, i.e. $k_\parallel = 0$. This is computationally advantageous, as it effectively reduces the problem to one dimension. Constructing a Lagrangian via L = $i\langle\psi|\partial_t|\psi\rangle - \langle\psi|H|\psi\rangle$, we deduce the corresponding nonlinear equations of motion:

$$\dot{\theta}_j = -S\left[J_{j-\frac{1}{2}}\cos\theta_{j-1}\sin(\varphi_j - \varphi_{j-1}) + J_{j+\frac{1}{2}}\cos\theta_{j+1}\sin(\varphi_j - \varphi_{j+1})\right.$$
$$+ 2\sin\varphi_j\left(D_j\sin\theta_j\right.$$
$$\left.\left. + K_{2,j}\cos\theta_j\cos\varphi_j - 2K_{4,j}\cos^3\theta_j\cos\varphi_j\sin^2\varphi_j\right)\right] \quad (17)$$

$$\dot{\varphi}_j = 2S\left[\left(4J_j - K_{2,j}\sin^2\varphi_j + 2K_{4,j}\cos^2\theta_j\sin^4\varphi_j\right)\sin\theta_j - D_j\frac{\cos(2\theta_j)}{\cos\theta_j}\cos\varphi_j\right.$$
$$+ \frac{J_{j-\frac{1}{2}}}{2}\left(\sin\theta_j\frac{\cos\theta_{j-1}}{\cos\theta_j}\cos(\varphi_j - \varphi_{j-1}) + \sin\theta_{j-1}\right)$$
$$\left.+ \frac{J_{j+\frac{1}{2}}}{2}\left(\sin\theta_j\frac{\cos\theta_{j+1}}{\cos\theta_j}\cos(\varphi_j - \varphi_{j+1}) + \sin\theta_{j+1}\right)\right] - \lambda\sin\varphi_j \quad (18)$$

where for each of the considered parameters $P \in \{J, D, K_2, K_4\}$ we denote by $P_j \equiv P(z = ja, t > 0)$ the quenched value, while the other three parameters are taken at their equilibrium value. Moreover, as discussed above (Extended Data Fig. 4), magnons at both $k = 0$ and $k \neq 0$ have finite, nearly identical relaxation rates in the absence of pumping. We incorporate this in our model by adding a phenomenological damping term, of strength $\lambda = 47.7 \ \mu\text{eV}$, to the Eq. (18) above.

## S3. Results of the modelling of the coherent magnon dynamics induced by effective quench of different magnetic interactions

### S3.1. The mechanism of the excitation of q-AFM mode

The generation of magnons following the optical pump can be understood as follows: altering the value of either interaction parameter $D$ or $J$ locally changes the equilibrium value of the canting angle, from $\theta_0$ to the new value:

$$\theta_r = \frac{1}{2}\arctan\left(\frac{D_r}{3J_r}\right) \quad (19)$$

Here, $D_r$ or respectively $J_r$ are the renormalized interaction strengths. Since in our model the spin state is not directly affected by the pump, this will give rise to precession around the new equilibrium, with amplitude $|\theta_r - \theta_0|$ (see Suppl. Fig. 3). Note that significant relative variation in $D$ or $J$ will only yield modest absolute changes of $\theta$, as the latter is small to start with ($D/3J \approx 1.75 \cdot 10^{-2}$). For example, a 67% reduction in $J$ is needed to raise the canting angle by only 1 degree. On the other hand, the azimuthal deflections attained during the evolution can be significantly larger, as the moments evolve along highly elongated trajectories.

In contrast, the anisotropy parameters $K_2$ and $K_4$ have no influence on the angle $\theta_0$. As a result, modifying them will not even suffice to start the q-AFM precession, what will be shown below in Supplementary S3.4.

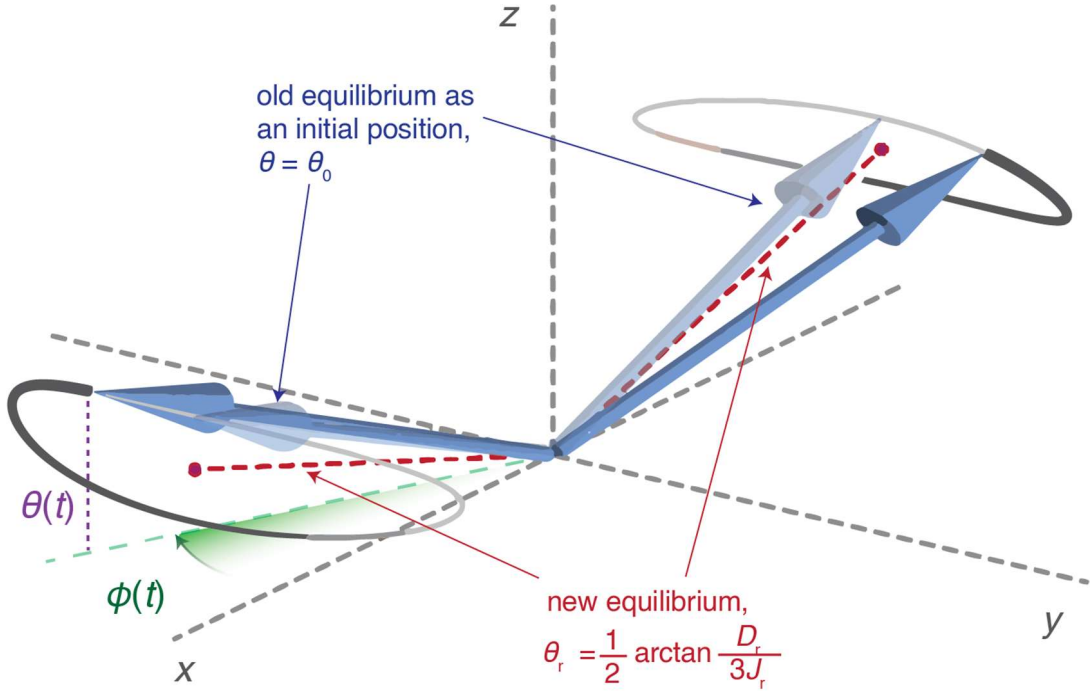

**Supplementary Figure 3: Quench of magnetic interactions as a driving mechanism for spin precession.** Semi-transparent blue arrows show spins in the equilibrium state with canting angle $\theta = \theta_0$. Red dashed lines indicate the new equilibrium state with canting angle $\theta_r$, produced by CT-induced renormalization of magnetic interactions to $J_r$ or $D_r$, thus starting the time-evolution of both canted angle $\theta(t)$ and azimuthal angle $\varphi(t)$.

*S3.2. Effect of localization on amplitude of the magnon dynamics at various momenta*

Under low-fluence pumping, when spin interactions are weakly renormalized (Suppl. Fig. 4a-d), the excitation of small-amplitude precession as described above is the only relevant effect. The initial amplitude of the precession will follow the same profile as the parameter quench,

$$|\theta_r - \theta_0|(z) \propto e^{-\frac{z}{\delta}} \quad (20)$$

whose Fourier transform will correspondingly have a profile centred at low $k$, up to a cutoff on the order of $\delta^{-1}$:

$$|\theta_r - \theta_0|(k) \propto \frac{1}{1+(k\delta)^2} \quad (21)$$

On the other hand, strong suppression of $D$ or $J$ will locally reduce the frequency of spin precession near the edge, allowing for responses below the equilibrium gap, and effectively creating a magnon trap. To quantify this, after the quench we examine the following magnetic potential:

$$H(z) = 2\sqrt{6}S \left[1 + \left(\frac{D(z)}{3J(z)}\right)^2\right]^{\frac{1}{4}} \sqrt{J(z)\left[K_2 + 3J(z)\left(\sqrt{1+\left(\frac{D(z)}{3J(z)}\right)^2} - 1\right)\right]} \quad (22)$$

This is the same expression as the Eq. (8) describing the magnon gap for the canted phase, but now the values of interaction parameters are understood to be the renormalized ones. Indeed, if the quench is weak and slowly-varying in space, $H(z)$ can be interpreted as a 'local gap'; under more abrupt changes, on the other hand, we employ it to quantify the depth of magnon trapping. Panels e-h of Suppl. Fig. 4 show the typical behaviour for strong renormalization, exemplified through a suppression of the exchange interaction $J$. The real-space plot (Suppl. Fig. 4h) displays a reduction in precession frequency near the surface, which gives rise to a visible mismatch with the corresponding one in the bulk. A relatively weaker wavefront still propagates inwards at the magnon velocity. The magnetic potential $H(z)$ is reduced near the surface, and in Suppl. Fig. 4g the spatially-resolved magnetization oscillations show an exponentially localized response, at frequencies below the equilibrium gap. The spatially integrated $m(k = 0, v)$ curve, shown on Suppl. Fig. 4f in red, displays the redshift and asymmetric broadening characteristic to the experimentally-measured signal.

Based on this picture, we can understand the anomalous behaviour of magnetization oscillations measured in the transmission geometry, its stark contrast to the $k > 0$ response seen in the reflection measurements, as well as the fluence dependence of the measured oscillation amplitudes (Suppl. Fig. 4i). Low-intensity pumping generates coherent wavepackets over a broad range of $k$, which afterwards evolve according to the original magnon dispersion. Increasing the drive strength in this regime will enhance the oscillation amplitude, yielding linear fluence dependence for the measured signal both at $k = 0$ and $k > 0$. In contrast, strong pumping creates a dip in the magnetic potential, which allows for response below the static magnon gap. The spatially integrated magnetization ($k = 0$) is sensitive to the nonequilibrium trapped state, and the respective signal will continue to grow with fluence. However, this further increase will happen at the expense of the $k > 0$ amplitude, which goes down as deeper trapping blocks most of the energy from escaping.

The wavepacket which does leave the pump's confinement region, and reaches the sample bulk, will experience unrenormalized interactions and in consequence the equilibrium dispersion again. The centre frequency and line shape of the main peak in finite−$k$ spectral response are not affected by the pump fluence (Extended Data Fig. 4), since they correspond to a wavepacket freely propagating in the bulk. Only at short times, when it crosses the pump confinement region, does the $k > 0$ packet also display anomalous response (Fig. 3a-b, main text).

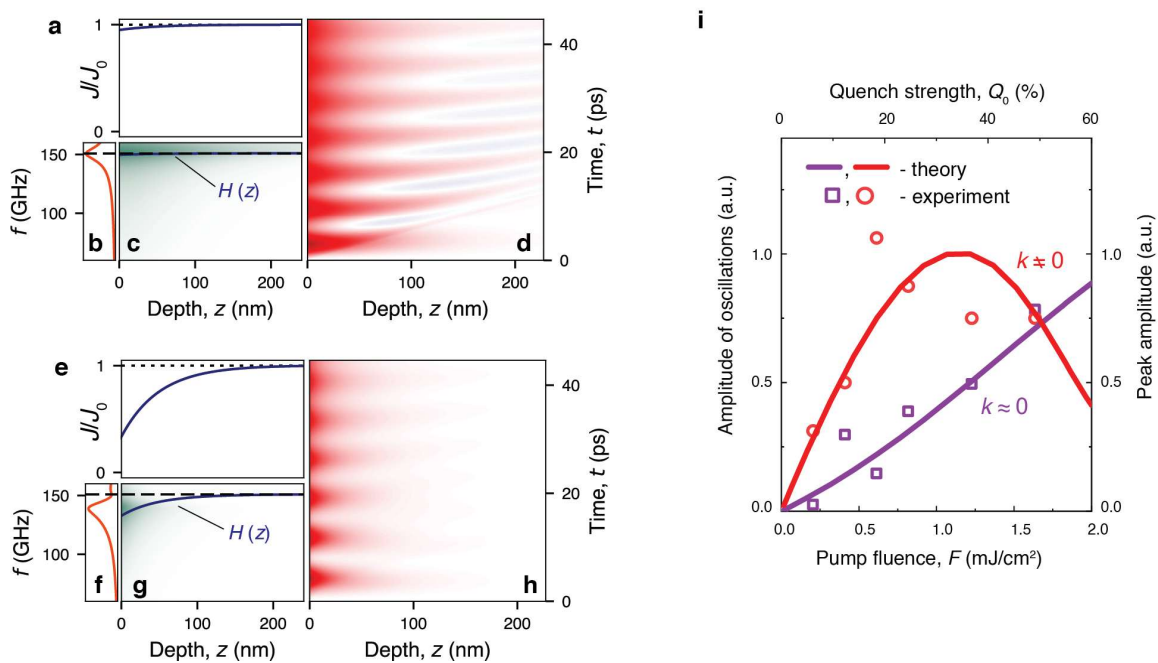

**Supplementary Figure 4: Quench-induced trapping of the magnon states.** Real-space picture of typical magnetic response under weak **(a-d)**, respectively strong (**e-h**) pumping. Panels **(a)** and **(e)** show the spatial profile of the quenched Hamiltonian parameter, in this case exchange interaction, exponentially recovering away from the sample edge. **(d)** and **(h)** illustrate the magnetization oscillations in real time, as well as the constant-velocity propagation of the magnetic wavefront at the limiting group velocity $V_\mathrm{m} = 2\sqrt{3}JSa$; while the connection between edge and bulk is smooth in the weak pumping case **(d)**, under strong driving **(h)** there is a visible frequency mismatch. Panels **(c)**, **(g)** overlay the calculated magnetic potential $H(z)$ (blue line) on the local intensity of frequency components in the magnetization (green shading). An exponentially-localized, in-gap response is visible in **(g)**. Finally, **(b)** and **(f)** respectively depict the spectrum of the $k \approx 0$ response, with anomalous behavior under strong driving. **(i)** Experimental values of the amplitude of oscillations as a function of pump fluence (symbols; left-bottom axes) plotted on top of the theoretical amplitude of the spectral peak as a function of the quench strength $Q_0$ (lines; right-top axes) for both of $k \approx 0$ (purple) and $k \neq 0$ (red) magnon states.

## S3.3. The effect of quenching of DM interaction D

In Supplementary S2.1 we discuss the connection between the photocarrier excitation and the exchange interaction $J$ for AFMs in the Hubbard model. Although, the similar mechanisms should be at play for DM interaction $D$, as the corresponding term in Eq. (1) couples the weak ferromagnetic moment to the Néel order. Kinetic frustration of the latter, in the vicinity of a mobile carrier, is therefore could be expected to suppress the effective DM interaction strength.

The results of the calculated coherent magnetic dynamics driven by quenching of $D$ is presented in Suppl. Fig. 5a. One can clearly see that with the suppression of $D$ also leads to the formation of the new band with the frequency below the magnon gap $\Delta$. Despite the reproducing the frequency shifted band, its bandwidth is significantly smaller than that in the experimental spectra, see Suppl. Fig. 5a, bottom panels.

Additionally, in contrast with $J$, the quench of $D$ does now correctly describe the behaviour of the amplitude of the localized dynamics at $k \approx 0$. Suppl. Fig. 5b shows that the amplitudes tend to be saturated and suppressed instead of nonlinear growth, what was seen in the experiment.

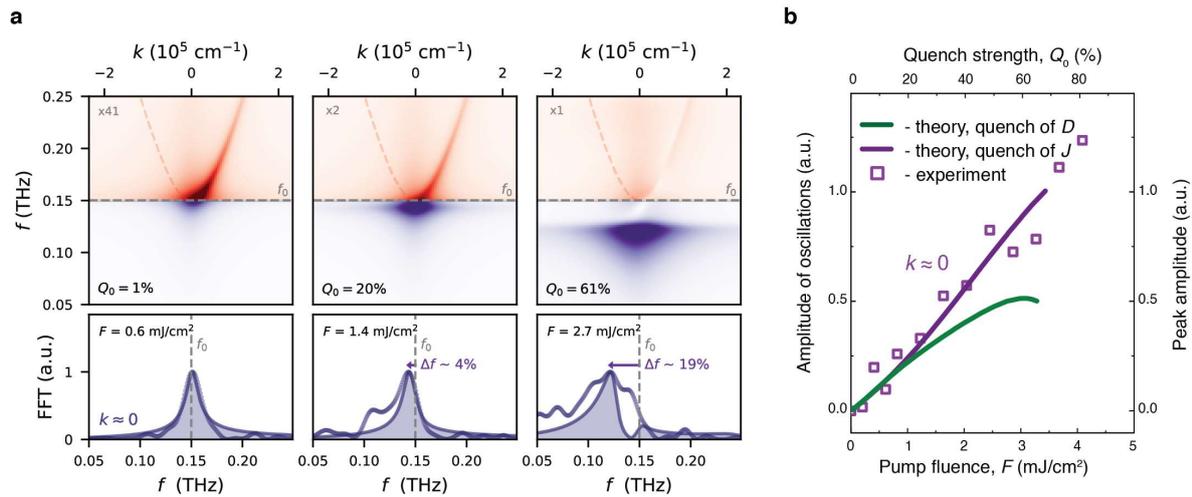

**Supplementary Figure 5: Ultrafast quench of DM interaction.** (a) (top panel) Reciprocal space snapshots of the numerically calculated coherent spin dynamics in the arbitrary scale (magnification is specified in the top-left corner) for different quenching strength $Q_0$. Red dashed line shows the equilibrium dispersion of spin waves. The straight grey dashed line shows the equilibrium magnon gap $f_0 = 0.15$ THz. (bottom panel) Corresponding linear cuts (solid lines with filled area) from the snapshots at $k = 0$ from the top panel mapped on top of the experimental Fourier spectra (dots) of the $k \approx 0$ dynamics presented in Fig. 2b. The straight grey dashed line shows the equilibrium magnon gap $f_0 = 0.15$ THz. The purple arrow highlights the excitation-induced red shift of the spectral peak $\Delta f$ which is observed in both of experiment and theory. The exact values of $Q_0$ were chosen for the best fit of the experimental data by the numerical calculations. **(b)** Experimental values of the amplitude of oscillations as a function of pump fluence (squares; left-bottom axes) plotted on top of the theoretical amplitude of the spectral peak as a function of the quench strength $Q_0$ (right-top axes) for quenching of either $J$ (solid purple line) or $D$ (solid green line) for $k \approx 0$ magnon states.

## S3.4. The effect of quenching of magnetic anisotropies $K_2$, $K_4$

Supplementary Figure 6 shows the resulting spectra after suppressing $K_2$ and $K_4$, with quench strengths comparable to the ones considered for $J$ and $D$. It is clearly seen that, even under strong quenching, the suppression of neither $K_2$ nor $K_4$ starts the spin dynamics. Therefore, the relevant contribution to the observed phenomena arises from $J$ and/or $D$.

Note that these observations are in an agreement with the presented mechanism of the excitation of the $q$-AFM mode, see Supplementary S3.1.

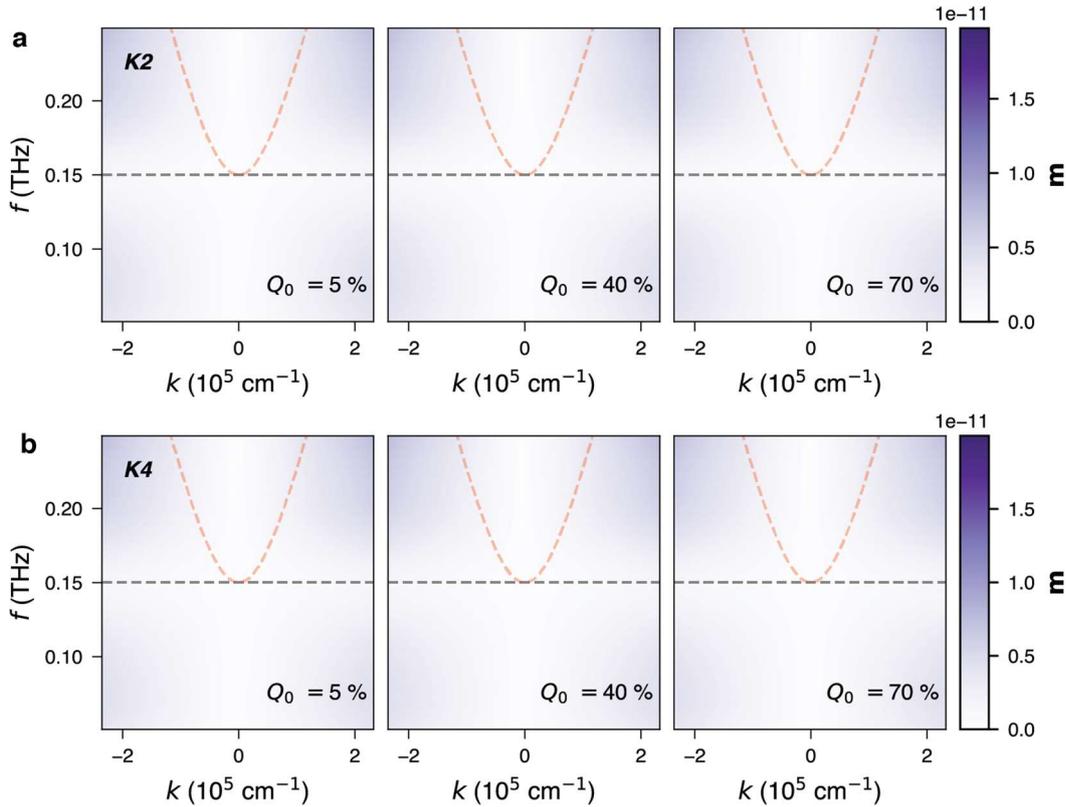

**Supplementary Figure 6: Modelling coherent magnon dynamics induced by ultrafast quench of 2$^{nd}$ and 4$^{th}$ order anisotropy.** Reciprocal space snapshots of the numerically calculated coherent spin dynamics, normalized by equilibrium magnetization, for low, medium and high quenching strength $Q_0$ for **(a)** the quadratic anisotropy $K_2$ quench and **(b)** quartic anisotropy $K_4$ quench. Red dashed line shows the equilibrium dispersion of spin waves. The straight grey dashed line shows a magnon gap $f_0$=0.15 THz.

*S3.5. The effect of reduction of local magnetic moments S*

An additional effect of pumping will be the spin pairing of electrons participating in charge excitations, thereby reducing the local moment size $\langle S^2 \rangle$. In our model, this can be implemented as a quench of the spin length $S$. We argue however that such an effect can be ignored, as it is relatively unimportant for understanding the current experimental results. Indeed, starting from $S = 5/2$, the pairing of two electrons will at most reduce the spin length to $3/2$, rather than suppress it completely. We may estimate for strong pumping that a 10% density of charge excitations, each reducing an on-site moment by 40%, yields an average decrease in $S$ of about 4%. As both the magnon gap (8) and limiting velocity (9) are linear in the moment size $S$, we expect all relevant energy scales to suffer reduction by at most the same fraction as $S$ does. This estimation is already at odds with the observed shift of spectral weight to much lower frequencies, as shown in Fig. 2 of the main text.

Supplementary Figure 7 demonstrates the results of the numerical calculations of the effect of suppressing $S$. Small reductions such as the above estimate $Q_0 = 4\%$ do not visibly alter the magnon spectrum, while much stronger quenching $Q_0 = 30\%$ yields an incorrect $k = 0$ lineshape, with spectral weight concentrated between the redshifted peak and the original magnon gap.

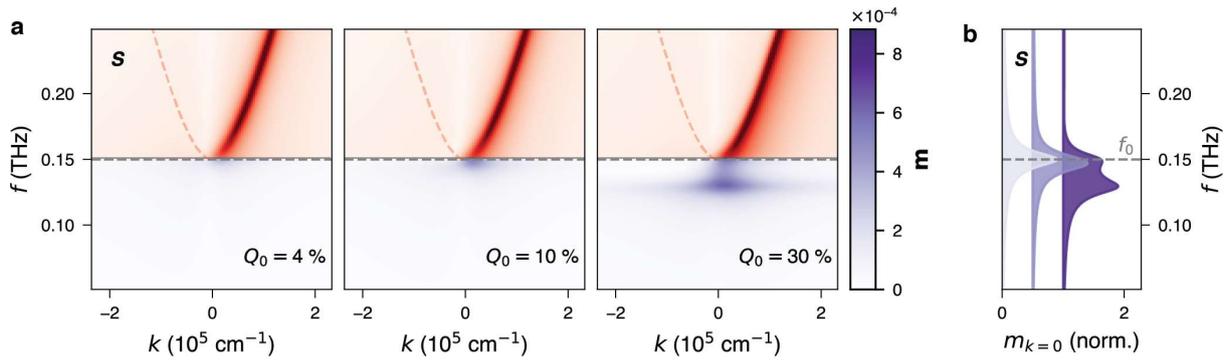

**Supplementary Figure 7: Modelling coherent magnon dynamics induced by ultrafast reduction of local magnetic moments $S$. (a)** Reciprocal space snapshots of the numerically calculated coherent spin dynamics, normalized by equilibrium magnetization, for low, medium and high quenching strength $Q_0$ for the length of spins $S$. Red dashed line shows the equilibrium dispersion of spin waves. The straight grey dashed line shows a magnon gap $f_0 = 0.15$ THz. **(b)** Renormalized spectra of $k \approx 0$ magnon mode induced by high quenching of $S$. The straight grey dashed line shows a magnon gap $f_0 = 0.15$ THz.